\documentclass[reprint,prc,amsmath,amssymb,superscriptaddress,noffotinbib,preprintnumbers]{revtex4-1}
\usepackage{url,color,colordvi,epsfig,pstricks}
\usepackage[colorlinks=true, citecolor=blue, filecolor=blue, linkcolor=blue, urlcolor=blue, pdftex]{hyperref}
\usepackage{graphicx,bm,dcolumn}
\usepackage{hyperref,slashed,lineno}
\usepackage{physics}
\usepackage{inputenc}
\usepackage{ulem}

\hypersetup{
    colorlinks=true, 
    linktoc=all, 
    linkcolor=blue, 
    citecolor=blue}

\begin{document}

\title{Modeling quasielastic interactions of monoenergetic kaon decay-at-rest neutrinos}
\author{A.~Nikolakopoulos}\email{alexis.nikolakopoulos@ugent.be}
\affiliation{Department of Physics and Astronomy, Ghent University, B-9000 Gent, Belgium}
\author{V.~Pandey}
\altaffiliation{Present Address: Fermi National Accelerator Laboratory, Batavia, Illinois 60510, USA}
\affiliation{Department of Physics, University of Florida, Gainesville, FL 32611, USA}
\author{J.~Spitz}
\affiliation{University of Michigan, Ann Arbor, MI, 48109, USA}
\author{N.~Jachowicz}\email{Natalie.Jachowicz@UGent.be}
\affiliation{Department of Physics and Astronomy, Ghent University, B-9000 Gent, Belgium}


\begin{abstract}

 Monoenergetic muon neutrinos at 236~MeV are readily produced in intense medium-energy proton facilities ($\gtrsim$2-3~GeV) when a positive kaon decays at rest (KDAR; $K^+ \rightarrow \mu^+ \nu_\mu$). These neutrinos provide a unique opportunity to both study the neutrino interaction and probe the nucleus with a monoenergetic weak-interaction-only tool. We present cross section calculations for quasielastic scattering of these 236~MeV neutrinos off $^{12}$C and $^{40}$Ar, paying special attention to low-energy aspects of the scattering process.  Our model takes the description of the nucleus in a mean-field (MF) approach as the starting point, where we solve Hartree-Fock (HF) equations using a Skyrme type nucleon-nucleon interaction. Thereby, we introduce long-range nuclear correlations by means of a continuum random phase approximation (CRPA) framework where we solve the CRPA equations using a Green's function method. The model successfully describes ($e,e'$) data on $^{12}$C and $^{40}$Ca in the kinematic region that overlaps with the KDAR $\nu_\mu$ phase space. In addition to these results, we present future prospects for precision KDAR cross section measurements and applications of our calculations in current and future experiments that will utilize these neutrinos.
 
\end{abstract}

\maketitle


\section{Introduction}

In the wake of the worldwide effort to disentangle the neutrino-oscillation puzzle, neutrino-nucleus interactions are now studied with unprecedented vigour and intensity.
The analysis of accelerator-based experiments requires a detailed modeling of the underlying neutrino-nucleus cross section~\cite{Alvarez-Ruso:2017oui}. As the neutrinos in these experiments are usually produced via charged particle decay-in-flight and hence arrive at the detector with a broad range of energies, a major stumbling block is the reconstruction of the incident neutrino energy in a particular interaction. In addition to issues associated with the uncertain reconstruction of the impinging particle's energy, determining the precise underlying reaction mechanism that gives rise to the experimental observables remains a cumbersome issue.

KDAR neutrinos offer a distinctive opportunity to study neutrino-nucleus interactions without having to deal with the complications raised by neutrinos originating from meson decay-in-flight with resulting broad energy-distributions.  These monoenergetic 236~MeV $\nu_\mu$ present an excellent tool for a more unambiguous calibration of cross sections and the determination of weak-interaction parameters, and carry the key to a better understanding of the role of, among others, initial and final-state interactions, correlations in the nuclear medium and multi-nucleon knockout. In general, KDAR $\nu_\mu$ can help to reduce experimental and theoretical uncertainties and ambiguities in an unprecedented way.

KDAR neutrinos are also highly relevant for understanding astrophysical neutrinos. In contrast to the typical GeV-scale neutrino sources produced by particle accelerators, especially from meson decay-in-flight, astrophysical neutrinos often have lower energies, usually in the 0-100~MeV region.  Still, their interactions with nuclei are, at best, just as poorly known as those of lab-made neutrinos. Owing to their relatively small energy, the 236~MeV KDAR $\nu_\mu$ therefore offers an important opportunity to gain insight into the interactions of astrophysical neutrinos.  These are especially important in the dynamics of type II supernovae and the nucleosynthesis processes taking place in these events. KDAR $\nu_\mu$ provide cross sections with energy transfers up to $E_{\nu}-m_{\mu} \approx 120$ MeV thereby covering a range of nuclear responses that is in the realm of supernova neutrinos. In particular, forward scattering of 236 MeV $\nu_\mu$ with low momentum transfers should be able to provide precious information about the low-energy excitations important in the interactions of astrophysical neutrinos, complementing experiments performed with neutrinos from muon decay-at-rest at lower energy (0-53~MeV)~\cite{Harada:2013yaa,oscsns,ccm, coherent_SMloi, Distel2003}.

In this work, we study the total and differential charged current quasielastic (CCQE) cross sections of 236~MeV KDAR neutrinos.  Importantly, the relevant energy transfer for these interactions is situated at the low-energy part of the genuine quasi-elastic region.  The impact of nuclear effects and low-energy excitations can therefore be expected to be relatively large, making a detailed microscopic modeling of these neutrino reactions,  with a careful treatment of nuclear physics effects, mandatory. In order to address these low-energy and nuclear-structure aspects with all due care, we use the continuum random phase approximation (CRPA) formalism discussed in Refs.~\cite{Ryckebusch:1988,Ryckebusch:1989,Jachowicz:1999,Jachowicz:2002,Jachowicz:2002,Jachowicz:2004,Jachowicz:2006,Pandey:2014,Pandey:2015,Pandey:2016,VanDessel:2018}.

The CRPA formalism starts from a Hartree-Fock (HF) model for the nuclear ground state. Wavefunctions for the outgoing nucleon are determined as a solution of the positive-energy Schr\"odinger equation fed with the same nuclear potential used for the calculation of the ground state. In this way, Fermi motion and binding are included naturally, and the influence of Pauli-blocking and elastic final state interactions are taken into account in a consistent way. While at high energy transfers the strong final state potential should be weaker than the HF potential of the initial state~\cite{RGJ:constraints}, at the low momentum transfers considered here the present approach should be a reliable approximation. The CRPA incorporates long-range correlations, indispensable for a proper description of the giant-resonance region, through a Green's function approach that results in a natural description of the single-particle continuum. The strength of the residual interaction is constrained using a dipole form factor with a cut-off mass fitted in a comparison with electron data~\cite{Pandey:2015}. Relativistic corrections are implemented following the procedure outlined in Ref.~\cite{Donnelly:1998,Pandey:2015}. The Coulomb corrections for the outgoing lepton in charged-current (CC) interactions are taken into account using the approach proposed in Ref.~\cite{Engel:1998}.
In order to take into account the finite width of the single-particle excitations, the nuclear responses are folded with a Lorentzian with a width of $3~\mathrm{MeV}$~\cite{Pandey:2015}. This folding procedure does not affect integrated cross sections, or the shape of the quasielastic peak. 
The effects of short-range correlations and meson-exchange currents, which where implemented in the same HF framework in Refs.~\cite{Tom:MEC, Tom:SRC}, are not included in the present results.

Experimentally, the KDAR neutrino has just begun to be studied, although there are now a number of existing and planned experiments that will be able to take advantage of it. As one example, the NuMI beamline absorber at Fermilab is a prominent KDAR $\nu_\mu$ source. A number of neutrino detectors in the vicinity of the NuMI beam absorber are sensitive to these neutrinos. Recently, MiniBooNE, a Cherenkov- and scintillation-based mineral oil detector located about 85~m from the NuMI beamline absorber, has successfully isolated 236~MeV KDAR $\nu_\mu$ events using muon energy reconstruction and timing information~\cite{Aguilar-Arevalo:2018ylq}. Thereby, the MiniBooNE collaboration reported the first (and only so far) measurement of monoenergetic KDAR $\nu_\mu$ interactions on $^{12}$C~\cite{Aguilar-Arevalo:2018ylq}. This result is discussed in more detail and compared to various generator- and model-based predictions below.

This paper is structured as follows: In Section~\ref{sec:kinematics}, we establish and discuss the kinematic region probed by CC scattering of a 236~MeV $\nu_\mu$ in terms of energy and momentum transferred to the nucleus. Then, in Section~\ref{sec:electron}, we compare the CRPA model to the electron scattering data that overlaps with this kinematic region.
Next, in Section~\ref{sec:neutrino}, the $\nu_\mu$-induced CC cross section is presented and studied in detail, and in Section~\ref{sec:future_xsec}, we present future prospects for precise KDAR cross section measurements, in particular on argon and carbon targets. Finally in Section~\ref{sec:experiments} we provide some insight in how modeling and measurements of the KDAR $\nu_\mu$ cross section could prove useful for neutrino oscillation and exotic-search experiments, both from direct measurements and by constraining the cross section itself.


\section{Kinematics}\label{sec:kinematics}
\begin{figure*}
\includegraphics{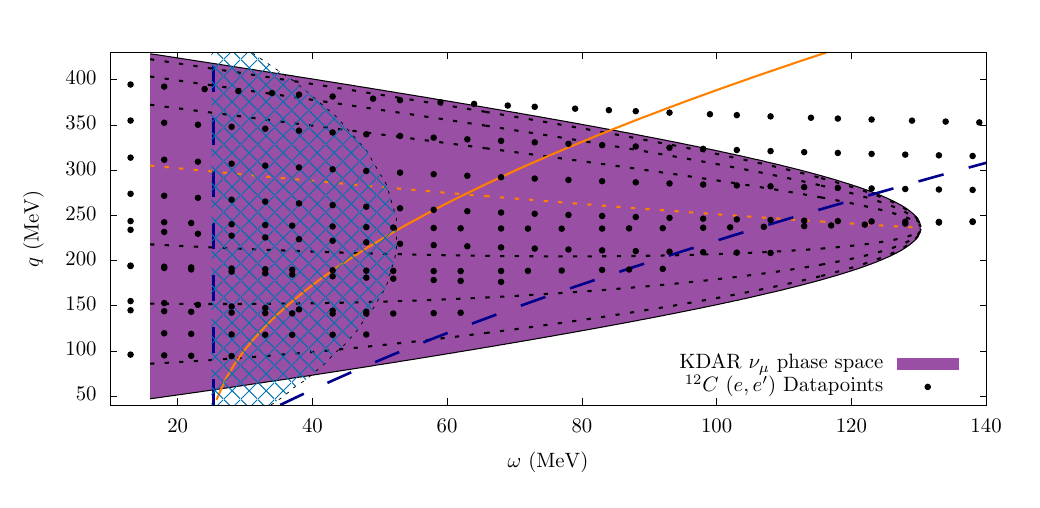}
\vspace{-.7cm}
  \caption{The phase space covered in $236~\mathrm{MeV}$ $\nu_\mu$ CC interactions is shown by the purple region. The dotted black lines denote fixed scattering angles, with the dotted orange line corresponding to $\cos\theta_\mu = 0$. In principle, this region extends to $\omega=0$, but as the lowest single-particle energy in carbon within the CRPA model is around 17~MeV, nucleons cannot be excited to the continuum below this threshold. The region in which the RFG response is non-zero lies between the blue dashed lines. The blue cross-shaded region shows the region that is affected by Pauli-blocking in the RFG. The available inclusive electron scattering data-points that overlap with the KDAR kinematic region are shown by the black dots. Finally, the full orange line shows the position of the quasielastic peak.}
\label{fig:phasespace}
\end{figure*}
The energy and momentum transferred to the nuclear system are defined from the lepton kinematics as
\begin{equation}
\omega = E_\nu - E_l,
\end{equation}
and
\begin{equation}
\label{eq:q}
q = \sqrt{E_\nu^2 + k_l^2 - 2E_\nu k_l \cos\theta_l},
\end{equation}
respectively. 
Here $k_l$ and $E_l$ are the charged lepton's energy and momentum, and $\theta_l$ is the scattering angle with respect to the incoming neutrino direction.
In terms of these variables, the available phase space for the CC interaction of a 236 MeV $\nu_\mu$ is given by the purple shaded region in Fig.~\ref{fig:phasespace}.
In principle, this region extends down to $\omega = 0$; the smallest energy transfer needed to excite a bound nucleon to the continuum in the CRPA model for ${}^{12}$C is shown. The contributions of discrete excitations of the nucleus are not considered in this work.
Interactions with fixed lepton scattering angles $\cos\theta_\mu$ run along the parabolae shown by the dotted lines. The dotted line denoting $\cos\theta_\mu = 0$ is shown in orange; $q$ values above  and below this line thus correspond to backward and forward scattered muons respectively.

\begin{figure}
\includegraphics[width=\columnwidth]{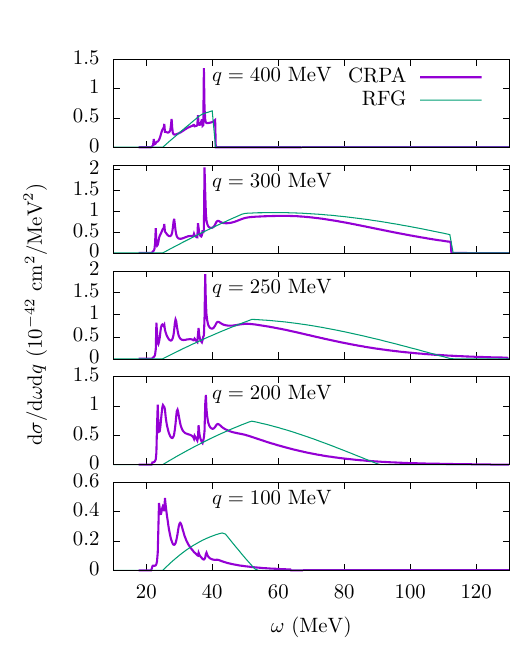}
\vspace{-.6cm}
  \caption{Double differential cross section for the CC interaction of a $236~\mathrm{MeV}$ $\nu_\mu$ with carbon for different values of $q$. }
 \label{fig:fixq}
\end{figure}

To appreciate the influence of low-energy nuclear effects on the scattering process it is instructive to contrast the CRPA with the relativistic Fermi gas (RFG) model.
In the RFG model, the interaction proceeds with a Fermi sea of non-interacting nucleons, occupying all momentum states up to the Fermi momentum $k_F$.
For carbon, we will use $k_F=228~\mathrm{MeV}$~\cite{Maieron:2002}.
Additionally, one introduces a binding energy $E_s$ in the RFG as a shift in the energy transfer by substituting $\omega \rightarrow \omega - E_s$ when computing the response.
We choose $E_s = 25~\mathrm{MeV}$, a value representing the average binding energy in the CRPA model for carbon.
The response in the RFG for certain values of $\omega$ and $q$ is proportional to the number of nucleons in the Fermi sea that can contribute to the interaction. 
The quasielastic peak is then the value for which this number is maximal. 
For a given energy transfer $\omega$, the peak position is given by
\begin{equation}
q^2_{QE} = \omega^2 + 2 M_N \left(\omega - E_s\right),
\end{equation}
which corresponds to the orange line in Fig.~\ref{fig:phasespace}.
In the RFG both the bound and outgoing nucleons are described by plane wave momentum eigenstates. 
Thus for a specific outgoing nucleon momentum and values of $\omega$ and $q$ only a single initial nucleon state can contribute.
This leads to the fact that the RFG response is only non-zero in between the dashed blue lines of Fig.~\ref{fig:phasespace},
as outside this region there are no nucleons available in the Fermi sea for an interaction to occur.

In the CRPA model by contrast, both the initial and final state nucleons occupy single-particle states which are energy eigenstates of a central potential.
The initial state nucleons have energies below the free nucleon mass. In carbon only 2  neutron shells are populated, leading to an average binding energy of around $25~\mathrm{MeV}$.
The threshold for a non-zero response to the continuum is not this average binding energy but rather the smallest single particle energy such that the available phase space goes down to lower $\omega$ than in the RFG.
The outgoing nucleon is also an energy eigenstate rather than a momentum eigenstate. Specifically the outgoing nucleon wavefunction behaves as an on-shell nucleon with fixed momentum for large distances from the nucleus, but is a mixture of momentum states in the nuclear interior.
This means that there is no sharp cut-off at high $\omega$ for a specific $q$ (or equivalently for high missing momentum) like in the RFG. Instead, the response smoothly goes to zero as determined by the high-momentum tail of the nuclear wavefunctions.

An important remark concerning the kinematics of the process involves the Pauli-blocked region. 
Again within the simple RFG model, owing to the orthogonality of the initial and final state plane waves, Pauli-blocking occurs if the outgoing nucleon's momentum is below $k_F$, in which case the interaction does not occur. 
The region in which this happens in the RFG is represented by the blue shaded region in Fig~\ref{fig:phasespace}.
From these considerations, it is clear that the KDAR phase space is situated in the transition region between the low-energy and genuine quasi-elastic regime, with large contributions from the threshold and Pauli-blocked regions, especially for forward lepton angles.

\begin{figure}
\includegraphics[width=\columnwidth]{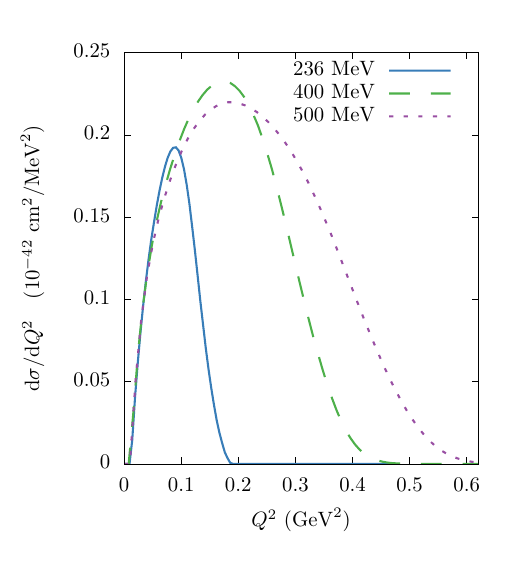}
\vspace{-1.cm}
  \caption{Differential cross section for the CC scattering of $\nu_\mu$ on carbon with different energies as a function of $Q^2$.}
 \label{fig:Q2}
\end{figure}
In the CRPA, Pauli-blocking is naturally included as all final-state wavefunctions are built from energy states that are obtained in the same mean-field potential as is used for the initial state.
Therefore the wavefunctions are orthogonal and thus satisfy the exclusion principle. To show the influence of Pauli-blocking on the cross section explicitly, the CRPA and RFG cross sections for fixed values of $q$ are shown in Fig.~\ref{fig:fixq}.
Here one clearly sees that the Pauli-blocked and threshold region account for a large part of the cross section strength, and that a simplified non-interacting model does not provide a satisfactory description.

It is also interesting to have a look at the $Q^2$ region covered in the interaction of a $236~\mathrm{MeV}$ $\nu_\mu$ CC scattering. This is shown in Fig.~\ref{fig:Q2}, along with the cross section for $500~\mathrm{MeV}$ $\nu_\mu$, an energy that lies close to the peak of the MiniBooNE~\cite{MB:flux} and T2K~\cite{T2K:flux} fluxes.
In recent years the low (reconstructed) $Q^2$ region has been the most challenging part of the phase-space to describe in MC generators. RPA calculations such as the ones of Nieves \textit{et al}.~\cite{Nieves:2004} have shown that including collective effects is paramount for describing the inclusive cross section in this kinematic region.
These calculations have found their way to the most popular neutrino event generators, albeit usually through an inclusive response table, while the hadronic side of the interaction is still taken from RFG-like momentum distributions. 
With the advent of measurements of the hadronic final state, it can be expected that a consistent treatment of inclusive and exclusive observables will become necessary, as evidenced by recent measurements from the MicroBooNE collaboration~\cite{Microboone:CCQE}.
In any case, it is clear that KDAR $\nu_\mu$ data can provide constraints on the cross section modeling in this low-$Q^2$ region.

Finally, we point out that there is significant overlap of the phase space available for CC scattering of $236~\mathrm{MeV}~\nu_\mu$ with a large body of inclusive electron scattering data. The datapoints for carbon are shown in Fig.~\ref{fig:phasespace} as an example. A combined analysis of this electron scattering and future KDAR $\nu_\mu$ data, with a model that treats the vector and axial currents consistently, could be revealing for the axial contribution to the nuclear current.

\section{Validation with electron scattering}\label{sec:electron}
\begin{figure*}
\includegraphics[width=\textwidth]{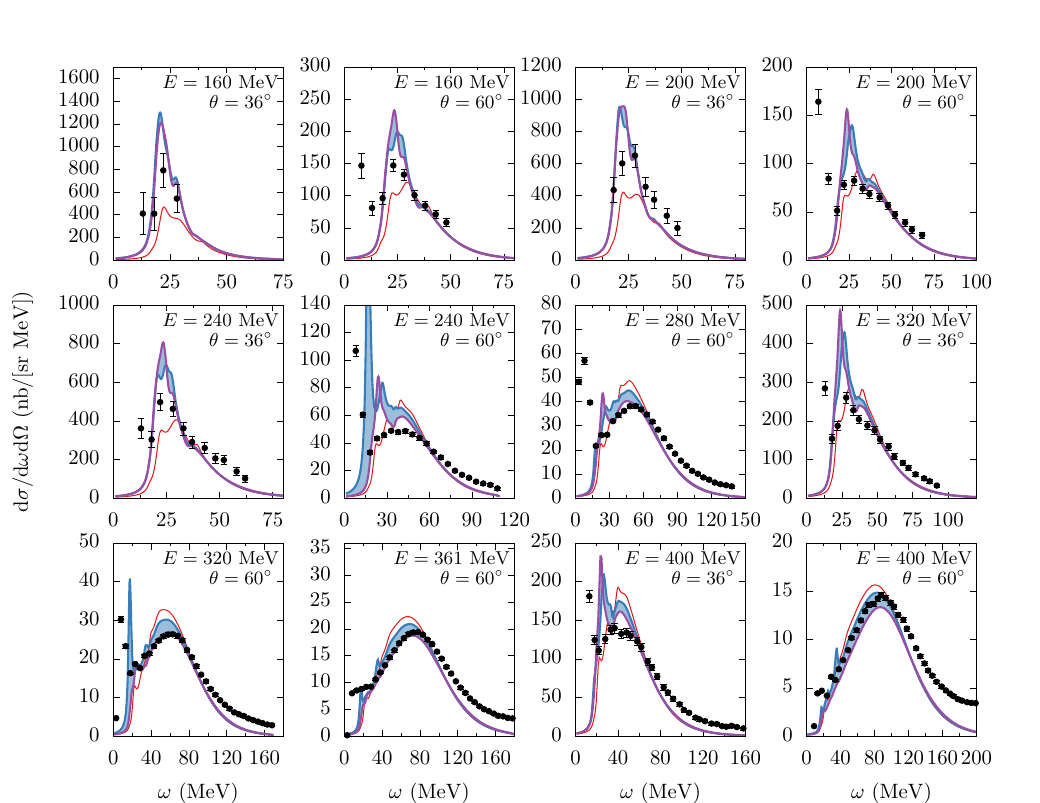}
  \caption{Comparison to ($e,e'$) data on ${}^{12}$C that overlaps in energy-momentum transfer with the $236~\mathrm{MeV}$ $\nu_\mu$ CC interaction. The Hartree-Fock results are plotted with the thin red lines, the CRPA cross sections are depicted with thick blue lines. The cross section where the dipole cut-off in the residual interaction is removed (see text) is shown in purple. The blue band illustrates the range in which the CRPA cross section can vary when the dipole cut-off mass is increased, yielding a smaller suppression of RPA effects at intermediate $q$.  
  The data points are from~\cite{Barreau:ee, QEarchive}.}
\label{fig:Eeprime_car}
\end{figure*}
\begin{figure*}
\includegraphics[width=\textwidth]{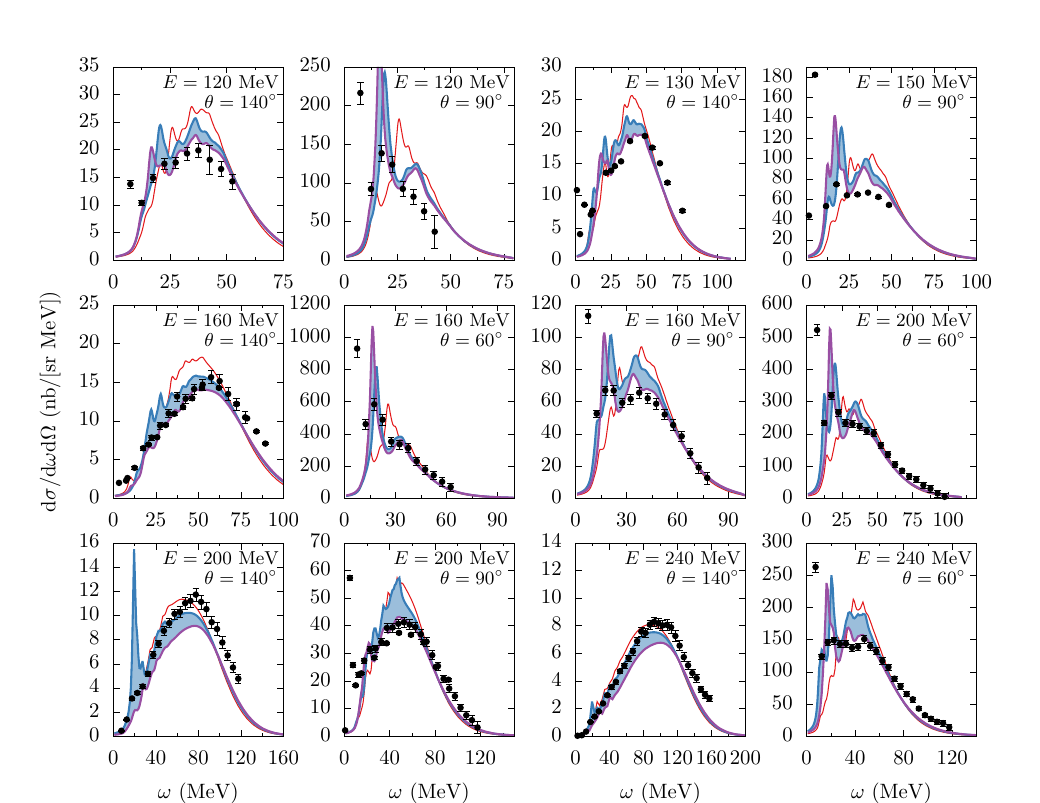}
  \caption{Comparison to ($e,e'$) data on ${}^{40}$Ca that overlaps in energy-momentum transfer with the $236~\mathrm{MeV}$ $\nu_\mu$ CC interaction. The labels are the same as in Fig.~\ref{fig:Eeprime_car}, The data points are from Refs.~\cite{Williamson:ee, Meziani:ee, QEarchive}. }
\label{fig:Eeprime_cal}
\end{figure*}

With the kinematics of the process established in the previous section, here we confront the CRPA model with the available inclusive electron scattering data on $^{12}$C and $^{40}$Ca that overlap with the same kinematic region in Figs.~\ref{fig:Eeprime_car} and ~\ref{fig:Eeprime_cal} respectively.
We point out that in the CRPA a dipole form factor is introduced in the residual interaction to regularize the zero-range Skyrme force at high values of $Q^2$.
The single parameter, a cut-off mass of $455~\mathrm{MeV}$, is determined from a global fit to $(e,e')$ data over a large kinematic range. 
As such, in this particular kinematic region where $q$ is comparable to the Fermi momentum, this suppression of CRPA effects tends to be too strong.
We therefore also show the result without the cut-off, and find a 5-10\% reduction at intermediate $q$, illustrated by the blue band, which can be considered as uncertainty.
In the rest of this paper we retain the dipole form-factor in the residual interaction consistent with~Ref.~\cite{Pandey:2015}

It is clear that a large part of the phase space relevant to the KDAR $\nu_\mu$ is sensitive to nuclear structure details, which are clearly shown in the CRPA cross sections.
The agreement of the CRPA model with the electron scattering data is reassuring for two reasons.
First, the charge changing weak vector current operator has the same structure as the electromagnetic current, as according to the conserved vector current (CVC) relation they are part of a conserved isotriplet, thereby directly relating the vector current of the weak interaction to those measured in electron scattering~\cite{Haxton:1995}.
This means that the ability to describe the inclusive electron scattering cross section data is a prerequisite for describing the vector current in neutrino interactions.

Second, a good agreement with the electron scattering data lends confidence in the nuclear model.
Indeed, we stress that the parameters of the CRPA model are not tuned in order to reproduce scattering data, except for the cut-off in the residual interaction.
The model relies on solving the Schr\"odinger equation for initial and final state nucleons in a nuclear potential.
This potential is obtained in a self-consistent Hartree-Fock calculation with an extended Skyrme force (SkE2), whose parameters are fit to reproduce static properties (charge radii, and excitation energies) of a representative set of closed-shell nuclei~\cite{Waroquier:Skyrme}. 
The agreement with electron scattering data is thus essentially a prediction that does not rely on a measurement of binding energies or momentum distributions.
This excellent agreement, especially in the low $\omega$ and $q$ region where the position of the single-particle resonances is correctly reproduced, thereby lends confidence to the predictive power of the model for the neutrino cross section in this kinematic region.

In confronting the model with both carbon and calcium data we find that the calcium data seem to be slightly better described by the CRPA.
On the one hand calcium is a larger, and moreover doubly magic, nucleus for which the mean-field treatment should be expected to work better.
On the other hand the calcium data is mostly obtained at backward scattering angles and at lower incoming energies than for carbon.
Thereby the data highlights a different weighting of the longitudinal and transverse responses that enter the cross section, with the backward scattering cross section having a larger contribution from the transverse response.

The simultaneous description of the different carbon and calcium data in this kinematic region without the modification of any parameters lends confidence to the ability of the model to provide predictions for different nuclei. 

\section{Neutrino scattering}\label{sec:neutrino}
In this section we present the cross section for CC neutrino-nucleus scattering in terms of the energy transfer and scattering angle.
This section is organized as follows, first in subsection~\ref{sec:kdarshape} we compare the predictions of the CRPA and several other models to the measurement of the muon energy distribution in the CC interaction of KDAR $\nu_\mu$ from MiniBooNE~\cite{Aguilar-Arevalo:2018ylq}.
Then in \ref{sec:Adep} we compare the cross section predictions for argon and carbon and explain how the A-dependence is handled in the CRPA model.
Next, in subsection~\ref{sec:responses}, the contributions of the different nuclear responses to the cross section are presented and compared to the cross section obtained at larger energies.
Finally, in subsection~\ref{sec:mpoles}, we show the contribution of different multipole moments to the cross section and illustrate that a consistent description of initial and final state wavefunctions has large effects on the description of the forward scattering cross section.

\subsection{Shape comparison to KDAR data\label{sec:kdarshape}}
\begin{figure}
\includegraphics[width=\columnwidth]{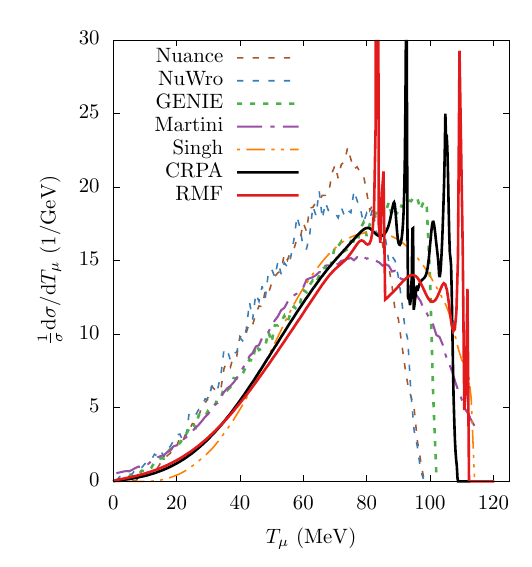}
\vspace{-.8cm}
  \caption{Comparison of the normalized (shape-only) KDAR $T_\mu$ distributions for various nuclear models.}
  \label{fig:models}
\end{figure}

\begin{figure*}
\includegraphics[width=\textwidth]{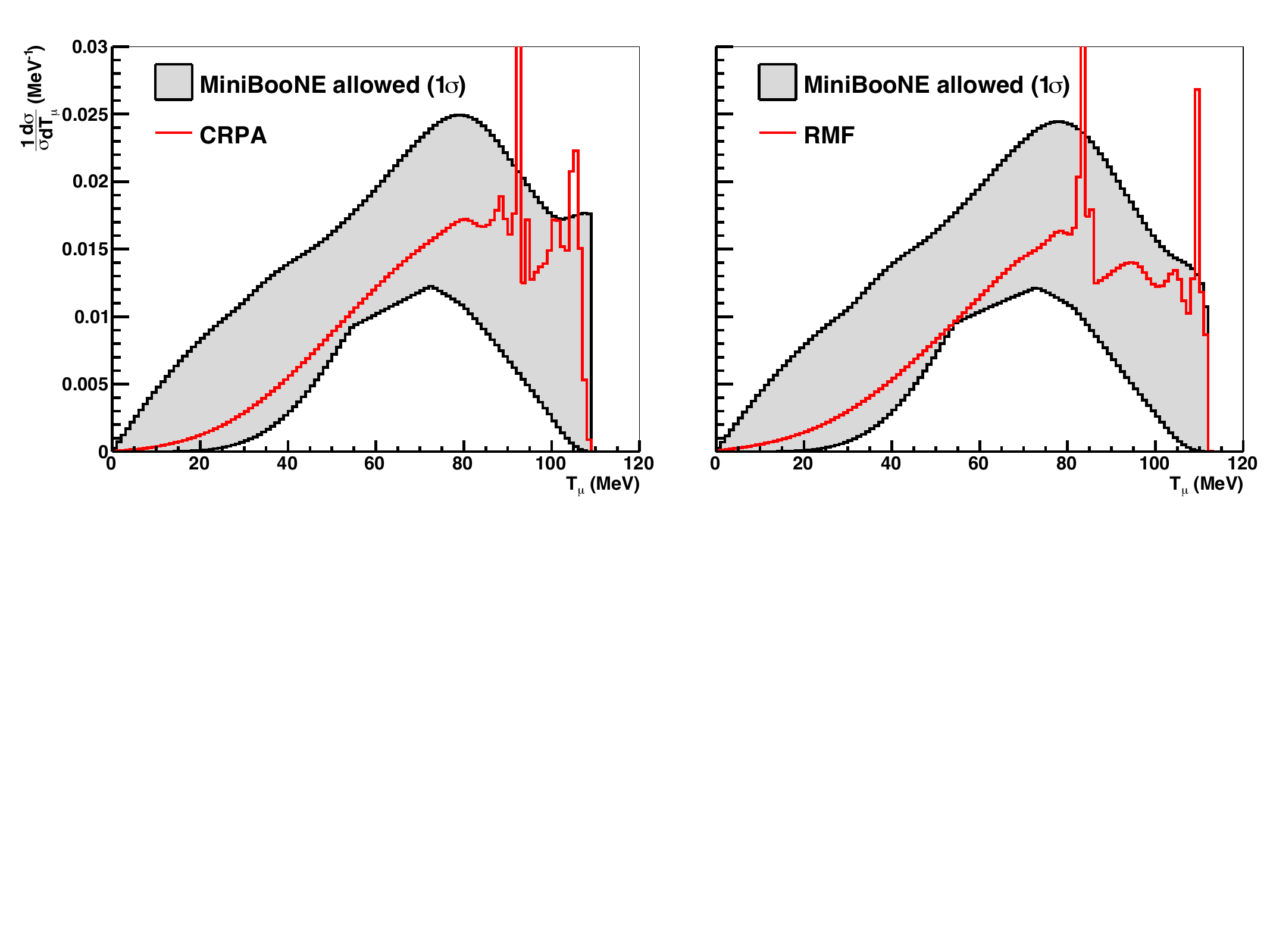}
\vspace{-6.8cm}
  \caption{A comparison between the shape-only MiniBooNE KDAR $T_\mu$ data (1$\sigma$ allowed region) and the CRPA (left) and RMF (right) predictions.}
 \label{fig:MB_comparison}
\end{figure*}

The MiniBooNE collaboration has performed a measurement of the shape-only distribution of muon energies for $\nu_\mu$ with $E_\nu = 236~\mathrm{MeV}$ on carbon~\cite{Aguilar-Arevalo:2018ylq}.
The data is available as an allowed region to which a shape-only comparison can be made with an online tool~\cite{KDARsite}.
In Fig.~\ref{fig:models} we show the $T_\mu$ distributions for several models available on the website. 
These include the results of the NuWro~\cite{NuWroKDAR}, Nuance~\cite{NuanceKDAR}, and GENIE~\cite{GENIEKDAR} event generators. Results are also given for the models of Martini \textit{et al}.~\cite{MartiniKDAR} and Singh \textit{et al}.~\cite{SinghKDAR}.
In addition to these, we show the results obtained within the CRPA model described here and the relativistic mean field (RMF) model of Ref.~\cite{RGJ:nucleareffects}. Table~\ref{tab:MB_comparison_table} shows how well the MiniBooNE data matches a number of predictions from various models and generators.
The value of $\chi^2_{prediction}$ is obtained from a direct comparison between the model in question and the MiniBooNE data, as described in Ref.~\cite{Aguilar-Arevalo:2018ylq}. The program takes the shape-only prediction (in 1~MeV bins), folds it into the detector observable, corrects for detector efficiency, and then performs a comparison to data to obtain the $\chi^2_{prediction}$. This is compared to the data best fit ($\chi^2_{\mathrm{best-fit}}$=72.6 with 64 degrees of freedom) to form a $\Delta \chi^2$ ($\chi^2_{\mathrm{prediction}}-\chi^2_{\mathrm{best-fit}}$). In the table the results obtained with the RFG model with a binding energy of $25~\mathrm{MeV}$ and $34~\mathrm{MeV}$ are also included and labeled RFG and RFG34 respectively.
To be complete, we also list the total cross sections for each model, while it should be understood that the $\Delta\chi^2$ is obtained in a shape-only comparison. 
Lastly, Figure~\ref{fig:MB_comparison} shows shape-only comparisons between the CRPA/RMF predictions and the data.

\begin{table}
      \begin{tabular}{|c|c|c||c|c|} \hline 
Model & $\Delta \chi^2$ & $\chi^2$ Prob. & $\sigma ~(10^{-39} \mathrm{cm}^2)$  \\ \hline \hline \hline
Nuance &  2.64 & 0.45 & 1.4  \\ \hline 
NuWro &  2.07 & 0.56 & 1.3 + 0.4 (np-nh)  \\ \hline 
GENIE &  0.95 & 0.81 & 1.75 \\ \hline 
Martini &  2.15 & 0.54 & 1.3 + 0.2 (np-nh)  \\ \hline 
Singh  &  3.90 & 0.27 & 0.91 \\ \hline 
CRPA  &  3.20 & 0.36 & 1.58 \\ \hline 
RMF &  3.49 & 0.27 & 1.56 \\ \hline
RFG &  1.69 & 0.64 & 1.66 \\ \hline 
RFG34 &  4.16 & 0.25 & 1.38  \\ \hline 
MB data &  - & - & $2.7\pm1.2$  \\ \hline 
\end{tabular} 
\caption{The compatibility of various models/generator predictions and the MiniBooNE data in terms of the shape-only KDAR $T_\mu$ (on carbon) spectrum. The comparisons are performed via the data release website~\cite{KDARsite}, and are reported in terms of $\Delta \chi^2$ ($\chi^2_{\mathrm{prediction}}-\chi^2_{\mathrm{best-fit}}$) for 3 parameters describing the shape and endpoint of the spectrum, and the associated probability. The data have a slight preference for some of the predictions over others, but the coarseness of the measurement doesn't allow any conclusions to be drawn.
Note that the total cross sections of the various models listed in the last column are included for reference and do not influence the $\Delta\chi^2$ comparison to data.}
  \label{tab:MB_comparison_table}
\end{table}

Due to the large uncertainty on the coarse MiniBooNE measurement, the data do not provide significant preferences among the different approaches.
Despite this, several remarks can be made regarding the results.
First, the endpoint for $T_\mu$ is an important parameter in the data, and different given endpoints provide a different shape.
This is because the location of the endpoint is not held fixed in the analysis, it is allowed to vary in the fit between 95-115~MeV, and the shape of the extracted spectrum is then dependent on this endpoint.
The high $T_\mu$ limit for QE scattering is determined by the minimal energy needed to eject a nucleon from the nucleus, which is not necessarily equal to the average binding energy of the nucleus as it is in e.g.~a RFG approach.
We see that the Nuance and NuWro results are similar to the RFG prediction with a binding energy of approximately $35~\mathrm{MeV}$.
The GENIE result shown is similar in shape, but with a lower threshold of around $25~\mathrm{MeV}$, and with a sharp fall-off around this threshold.
The other models have a lower threshold, and are rather similar in shape.
The data tends to prefer models with a larger threshold; in reality, however, the CC cross section could extend down to nearly zero energy transfer when discrete excitations of the nucleus are taken into account.

The Martini and Singh models are similar in shape, with the main difference between them that the Martini model provides more strength in the low-$T_\mu$ tail, while the Singh model places this strength in the high-$T_{\mu}$ region.
Note, however, that the latter predicts a significantly smaller total cross section than the other models because of a strong quenching due to RPA corrections~\cite{SinghKDAR}.
The CRPA and RMF models place even more strength in the high $T_\mu$ peak compared to the tail due to single particle resonances that arise in the mean field.
These effects are more apparent for the CRPA results, which include collective strength induced by long-range effects explicitly.

There are other approaches not included in this comparison which could provide a description of CC scattering off carbon in this kinematic region. These include the aforementioned RPA model of Nieves et~al.~\cite{Nieves:2004}, and several spectral function approaches that include FSI~\cite{Rocco:PRC100,Rocco:EFS,Ankowski:FSI}. Recently, electroweak responses of carbon for $100~\mathrm{MeV} \leq q \leq 700~\mathrm{MeV}$ were computed ab initio using quantum Monte Carlo methods with a consistent treatment of one- and two-body currents~\cite{Lovato:MBT2K}.

Although there are significant micro- and macro-shape differences between the various models in terms of $T_\mu$/$\omega$, we note that the experimental ability to reconstruct muon angle, in addition to $T_\mu$, and therefore $q$ and $Q^2$, would allow differentiation between models more readily, as demonstrated in Figs.~\ref{fig:phasespace} and \ref{fig:fixq}. At $q$-values of 200~MeV or less, for example, significant differences in the shape of the RFG and CRPA cross sections are seen, and the cross section becomes quite sensitive to the treatment of Pauli-blocking and binding energy. However, even with $T_\mu$ only and although the single-MeV-scale width nuclear resonances predicted by CRPA and RMF are likely difficult to resolve, the significant differences between peak position and binding energy, on the order of 20~MeV, among the models should be readily resolvable by all the experiments discussed in Sec.~\ref{sec:future_xsec}, even those without the ability to measure muon angle. Further, the ability to reconstruct the hadronic component of the interaction also provides additional power for model comparisons as well. Prospects for KDAR measurements in terms of a number of experimental observables with various detector technologies are discussed below in Sec.~\ref{sec:future_xsec}.

\subsection{Coulomb effects and A-dependence \label{sec:Adep}}
First, we describe how the effect of the Coulomb potential of the nucleus on the charged lepton is taken into account.
In principle this problem can be solved by treating the outgoing lepton as an energy eigenstate in the Coulomb potential of the nucleus.
In general Coulomb effects are small for leptons with energies of a couple hundred MeV and for nuclei lighter than iron. They can be estimated quite adequately for integrated cross sections by using the modified effective momentum approximation (MEMA)~\cite{Engel:1998}.

In the MEMA approach the outgoing lepton plane wave's momentum and normalization is modified as
\begin{equation}
e^{i\vec{k}\cdot\vec{r}} \rightarrow \sqrt{\frac{E_{eff}k_{eff}}{Ek}}e^{i \vec{k}_{eff}\cdot\vec{r}}
\end{equation}
with $k_{eff} = k - V(0)$, and $E_{eff} = \sqrt{k_{eff}^2 + m_l^2}$, with $m_l$ the charged leptons mass.
In this work the potential $V(r)$ corresponds to the Coulomb potential of a spherical nucleus with charge $Z^\prime = Z + 1$ and radius approximated as $R=1.2A^{1/3}~\mathrm{fm}$.
This thus introduces a fixed shift in the momentum transfer entering the matrix element by an amount $\frac{3}{2}Z^\prime\alpha/R$ and an overall phase factor $\frac{k_{eff}E_{eff}}{k_l E_l}$ in the cross section. 

\begin{figure}
\includegraphics[width=\columnwidth]{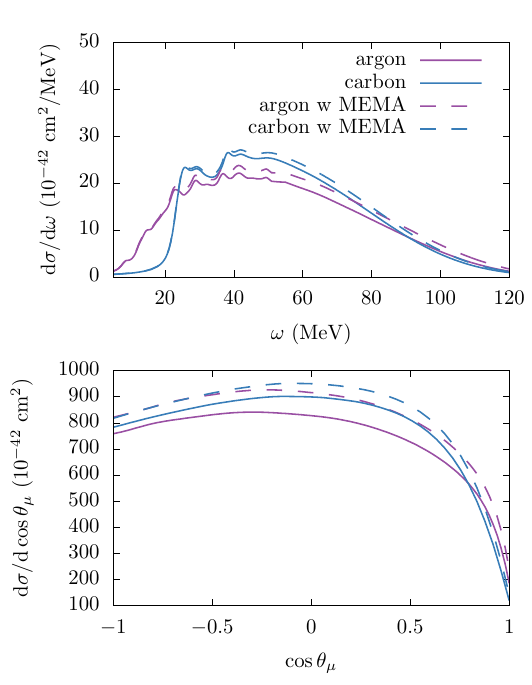}
\vspace{-.6cm}
  \caption{Cross sections per neutron as a function of the energy transfer (top panel) and muon scattering angles (bottom panel) for the CC interaction of the $236~\mathrm{MeV}$ $\nu_\mu$ with carbon and argon. The dashed lines are CRPA predictions with MEMA corrections.}
  \label{fig:Carbon_argon}
\end{figure}

For small outgoing energies the correction from MEMA becomes inappropriate and the effect of the Coulomb distortion can be included by making use of the Fermi factor, which is the ratio of a plane wave and the Coulomb wave function if only s-wave contributions are taken into account~\cite{Engel:1998}. 
In our calculations, we interpolate between both approaches as detailed in Ref.~\cite{VanDessel:Forbidden}, but for muon cross sections in this work this procedure naturally leads to the MEMA correction.
The effect of the MEMA for $\nu_\mu$ scattering is shown explicitly in Fig.~\ref{fig:Carbon_argon} where results for CC scattering off argon and carbon are computed with and without the MEMA.
The main increase stems from larger energy transfers, i.e. smaller outgoing muon energies, as expected. 
We apply the appropriate Coulomb corrections for all further results in this work unless mentioned otherwise. 

Carbon, oxygen, and increasingly argon are the predominant targets in current and future neutrino scattering experiments. It is therefore important to study variations of the cross section across a range of nuclear masses. 
Cross sections for neutrino interactions with different nuclei computed within the CRPA approach were presented previously in Ref.~\cite{VanDessel:2018}.
Here we restrict ourselves to interactions with carbon and argon as these are the targets used in experiments that aim at measuring KDAR cross sections.
In Fig.~\ref{fig:Carbon_argon} we first look at the single differential cross sections in terms of energy transfer and scattering angle.
The cross sections for CC scattering of carbon and argon are shown, normalized per neutron.
The main complication for calculations on argon lies in the fact  that the outerlying proton and neutron shells are not fully occupied. 
We therefore compute the argon single particle wavefunctions and mean-field potential with the same self-consistent Hartree-Fock procedure as for carbon and calcium, but with fractional occupation numbers for these outerlying shells.
This procedure yields good results for the charge-distribution and compares well with RMF~\cite{Yang:coherent} and no-core shell-model~\cite{Payne:coherent} calculations for the neutron distributions~\cite{VanDessel:coherent}.
In Ref.~\cite{RGJ:constraints}  good agreement was found between the RMF and HF/CRPA cross sections for electron scattering off both argon and carbon.

An important difference with respect to the KDAR measurements is the significantly smaller threshold for single nucleon knockout in argon compared to carbon. 
The forward scattering region mostly overlaps with the peak in the low-$\omega$ region at small $q$, which means that the argon cross section becomes larger for forward scattering.
Also, we note that the increase of the cross section due to the Coulomb potential is naturally larger in argon as the nuclear charge is bigger.

\subsection{Separation of lepton and hadron vertices: nuclear responses\label{sec:responses}}
\begin{figure*}
    \centering
    \includegraphics{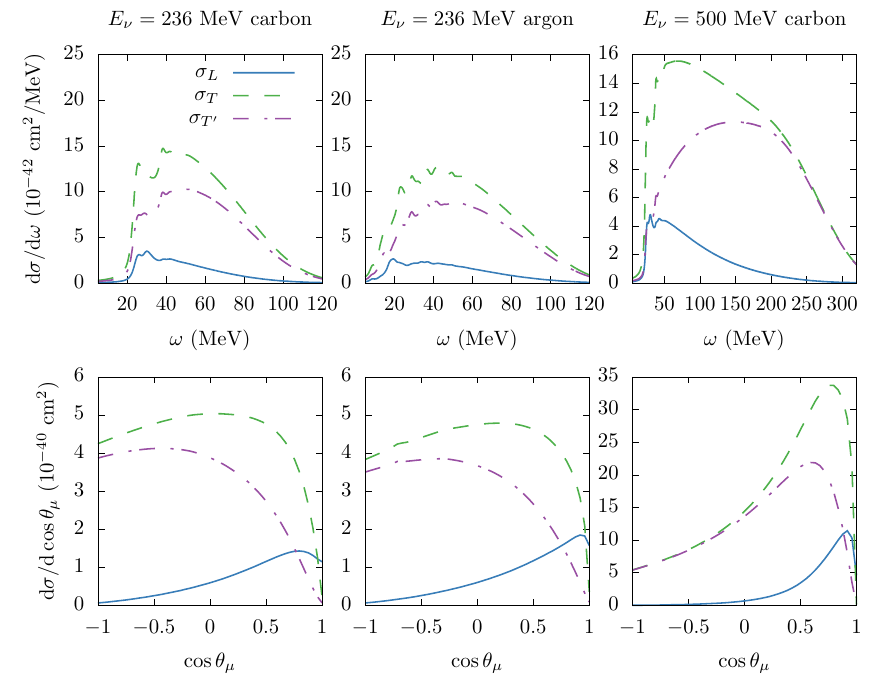}
    \vspace{-.3cm}
    \caption{Longitudinal and transverse contributions to the cross section per neutron for CCQE scattering of $\nu_\mu$ on carbon and argon for different incoming energies.}
    \label{fig:Responses}
\end{figure*}

It is common and instructive to write the cross section for leptonic scattering off nuclei in terms of specific factors that arise from the lepton and hadron vertices. In this way the dependence on the lepton kinematics is made clear. The inclusive nuclear response functions that arise in this separation then only depend on the energy and momentum carried by the exchanged gauge boson.
We only give the expressions necessary to define the results presented in the following sections; for a complete derivation of these expressions we refer the reader to Refs.~\cite{Moreno:2014, Walecka:book}.

The cross section for a lepton with four momentum $k_i$ scattering off a nucleus where a single lepton with four momentum $k_f$ is detected, assuming a single gauge boson is exchanged between the leptonic and hadronic vertex, can be written as
\begin{align}
\frac{\mathrm{d}\sigma_X}{\mathrm{d}E_f\mathrm{d}\cos\theta_f}&=\frac{F_X^2 E_f k_f}{2\pi}\times  \nonumber\\ 
& \left[ \left(V_{CC}R_{CC} + V_{CL}R_{CL} + V_{LL}R_{LL} \right)\right. \label{eq:sigmal} \\
& +\left.\left( V_{T} R_T + hV_{T^\prime}R_{T^\prime} \label{eq:sigmat}\right)\right].
\end{align}
Here the coupling $F_X^2$ depends on the interaction and is $\frac{e^4}{4Q^2}$ or $\frac{G_F^2}{2}\cos^2\theta_c$ for electromagnetic and weak interactions, respectively.

When the hadronic final state remains completely undetected the only relevant direction for the nuclear system is that of the momentum transfer $\vec{q}$.
This means that no mixing of longitudinal and transverse terms relative to the direction of the momentum transfer can occur and  the cross section is the sum of the longitudinal and transverse parts shown in Eqs.~\ref{eq:sigmal} and~\ref{eq:sigmat} respectively.
The last term vanishes when averaged over the helicity $h$ of the incoming lepton, as is the case in unpolarized electron scattering, but is always present for neutrino scattering.  We will denote it by $\sigma_{T^\prime}$, while the first term in Eq.~\ref{eq:sigmat} is denoted as $\sigma_T$.

We stress that the nuclear responses only depend on the energy and momentum transfer $\omega$ and $q$, and not directly on the lepton kinematics.
The dependencies of the cross section on the lepton kinematics are completely determined by the $V_i$ factors which are easily evaluated for plane wave leptons, resulting in~\cite{Walecka:book, Moreno:2014, Donnelly:1985}
\begin{align}
V_{CC} &= 1 + \frac{k_l}{E_l}\cos\theta_l, \nonumber \\
V_{CL} &= -\left(\frac{\omega}{q}V_{CC} + \frac{m_l^2}{E_lq}\right), \nonumber\\
V_{LL} &= V_{CC} - \frac{2E_iE_l}{q^2}\left(\frac{k_l}{E_l}\right)^2 \sin^2\theta_l, \label{eq:Vi} \\
V_{T} &= 2 - V_{CC} +  \frac{E_iE_l}{q^2}\left(\frac{k_l}{E_l}\right)^2 \sin^2\theta_l,\nonumber \\
V_{T^\prime} &= \frac{E_i+E_l}{q}\left(2-V_{CC}\right) -\frac{m_l^2}{E_f q}. \nonumber 
\end{align}
The nuclear response functions will be further described in Sec. \ref{sec:mpoles}, where we consider the multipole decomposition. 
Here we will just add that all the different terms in the cross section of Eqs.~\ref{eq:sigmal} and~\ref{eq:sigmat} are sums of purely vector-vector and axial-axial contributions, except for $\sigma_{T^\prime}$ which is a vector-axial interference term.

\begin{figure*}
    \centering
    \includegraphics{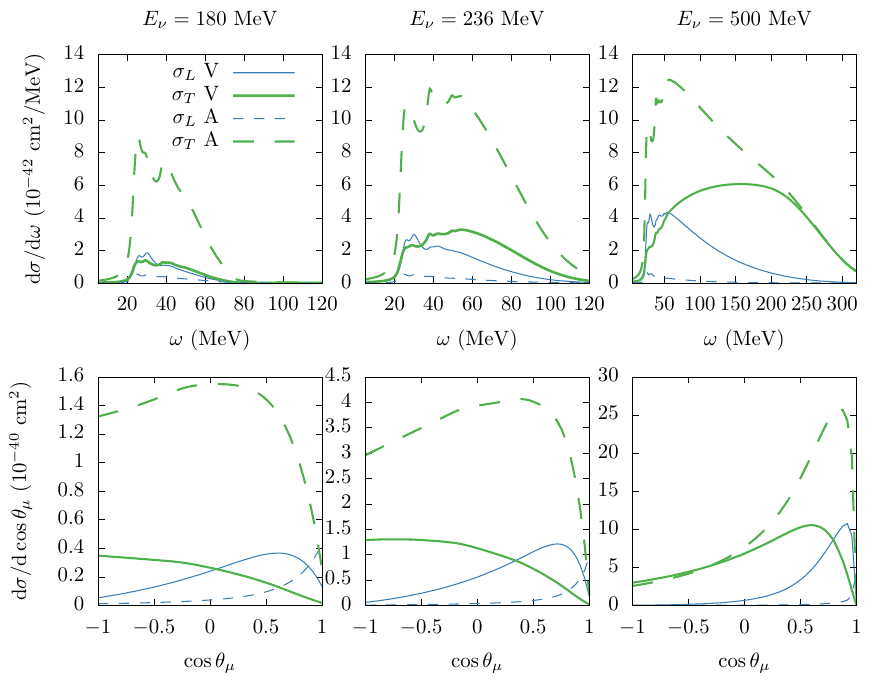}
    \vspace{-.3cm}
    \caption{ Longitudinal and transverse contributions to the cross section per neutron for CCQE scattering of $\nu_\mu$ on carbon for different incoming energies, separated in terms of the vector (solid lines) and axial (dashed lines) contributions.}
    \label{fig:ResponsesVA_C}
\end{figure*}

In Fig.~\ref{fig:Responses} we show the cross section for CC scattering separated into the longitudinal and both transverse contributions.
We show cross sections for $E_\nu = 236~\mathrm{MeV}$ for both carbon and argon. Additionally the cross section for $E_\nu = 500~\mathrm{MeV}$, closer to the peak  of the MiniBooNE and T2K neutrino fluxes, is shown in order to highlight the differences.
It is clear that the KDAR cross sections are far more sensitive to the nuclear structure and  associated single-particle excitations than the $500$ MeV cross section whose main strength stems from higher energy transfers.
For 236~MeV $\nu_\mu$, the dominant contribution is provided by the transverse response, especially at backward scattering angles, while the forward peak seen for the higher incoming neutrino energies disappears due to Pauli-blocking.
In Fig.~\ref{fig:ResponsesVA_C}, the longitudinal and transverse cross sections are shown separated into their vector and axial contributions.
This shows that the longitudinal cross section is dominated by the vector current, while the transverse cross section is mostly axial.
The small axial contribution in the longitudinal cross section is forward peaked, and grows in relative importance for smaller incoming energies.
If the lepton mass can be neglected, the vector contribution to the longitudinal current becomes exactly zero for completely back and forward scattered leptons. 
If the lepton mass cannot be neglected, the cross section for backward angles is a mix between a small longitudinal and a larger transverse contribution.  From the different panels, it becomes clear that the relative weight of various contributions to the total interaction strength can be quite different for 236~MeV  and higher energy neutrinos.  In general, lower energy cross sections have large contributions from the axial current, and feature a strong promotion of backward scattering processes.

\subsection{Multipole decomposition\label{sec:mpoles}}
\begin{figure}
    \centering
    \includegraphics{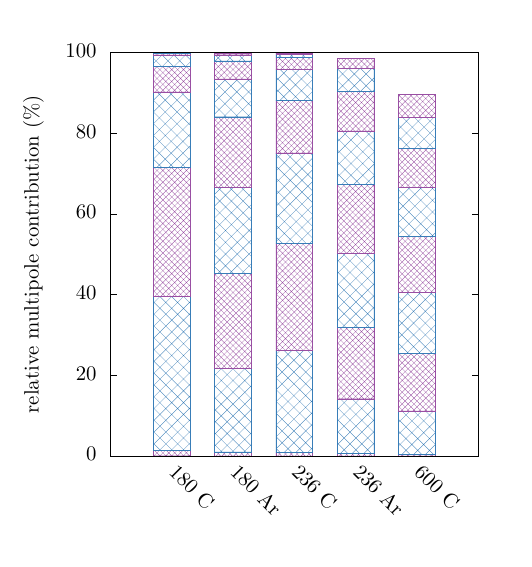}
    \vspace{-1.2cm}
    \caption{Relative contributions of multipoles 0-8 to the total cross section for different incoming energies (in MeV) and targets in a stacked histogram. The bottom bar corresponds to $J=0$ and $J$ increases going upwards.}
    \label{fig:Mpoles}
\end{figure}

Here we take a closer look at the nuclear response functions introduced in the previous section and show how a multipole decomposition is performed. 
By separating the responses in multipoles of total angular momentum of the exchanged boson one gets additional insight into the dynamics of the problem.
Different multipole moments not only correspond to a different weighting of the scattered lepton's angle but also of the angular momentum of the hadronic final state.
 
 Again we only give the essential expressions to illustrate the results presented in this section; for a complete derivation we refer the reader to Refs.~\cite{Walecka:book, Donnelly:1985}.
The expression for the nuclear current in terms of a general transition operator $\mathcal{O}^\mu$ is
\begin{eqnarray}
\label{eq:Jmu}
J^\mu\left( q\right) &=& \int d\vec{r} e^{-i\vec{q}\cdot\vec{r}}  \Psi^*_f\left(\vec{r}\right) \mathcal{O}^\mu\left( \vec{r} \right) \Psi_i\left(\vec{r}\right) \\&=& \int d\vec{r}e^{-i\vec{q}\cdot\vec{r}} \mathcal{J}^\mu\left(\mathbf{r}\right),
\end{eqnarray}
where $\Psi_i$ and $\Psi_f$ are the wavefunctions describing the initial and final nuclear state, respectively.
As we work in a spherically symmetric potential, the initial state is described by a Slater determinant of single particle angular momentum eigenstates.
The single particle wave function is described by the product of a radial wavefunction and a spin-spherical harmonic $\ket{ljm}$ as,
\begin{align*}
\phi_{jlm}\left(\vec{r}\right) &= \psi_{lj}\left(r\right)\sum_{m_l,m_s}\braket{l~m_l~1/2~m_s}{j m}Y_{l m_l}\left(\Omega_r\right)\chi_{m_s} \\
                                      &= \psi_{lj}\left(r\right)\ket{ljm}.
\end{align*}
This is also the case for the final-state wave functions. In a central potential, the  wave function for an outgoing nucleon with asymptotic momentum $\mathbf{p}_N$ can be written as a sum of partial waves that are angular momentum eigenstates.
With the single particle wave functions written as spin-spherical harmonics the only remaining ingredient for the partial wave expansion of the matrix element is the phase factor $e^{i\mathbf{r}\cdot\mathbf{q}}$.
For this one we make use of the partial wave expansion of a plane wave in terms of spherical Bessel functions and write
\begin{align*}
e^{i\vec{r}\cdot\vec{q}} &= 4\pi \sum_{JM} i^J j_J\left(qr\right)Y^*_{JM}\left(\Omega_r\right)Y_{JM}\left(\Omega_q \right) \\
&= \sqrt{4\pi}\sum_{J}\sqrt{2J+1}i^Jj_J\left(qr\right)Y_{J0}\left(\Omega_r\right),
\end{align*}
where in the second equality we have explicitly oriented the coordinate system with $\vec{q}$ along the $z$-axis such that only $M=0$ contributes.

It is now instructive to decompose the current of Eq.~(\ref{eq:Jmu}) into its spherical components in the $\vec{q}$ along z system, and to project this vector function onto irreducible tensor operators.
This is done by making use of vector spherical harmonics $\vec{\mathcal{Y}}^{M}_{J(L,1)}$; the necessary operators are defined e.g. in Refs~\cite{Donnelly:1985,Walecka:book} and yield the following form for the irreducible tensor operators
\begin{align}
\mathcal{M}_{JM} &= \int\,d\vec{r} \left[j_{J}\left(qr\right)Y_{JM}\left(\Omega_r\right)\right]\mathcal{J}^0\left(\vec{r} \right) \label{eq:M} \\
\mathcal{L}_{JM} &= \frac{i}{q}\int\,d\vec{r} \left[\vec{\nabla}\left(j_{J}\left(qr\right)Y_{JM}\left(\Omega_r\right)\right)\right]\cdot\vec{\mathcal{J}}\left(\vec{r} \right) \\
\mathcal{T}^{el}_{JM} &= \frac{1}{q}\int\,d\vec{r} \left[\vec{\nabla}\cross j_{J}\left(qr\right)\vec{\mathcal{Y}}^{M}_{J(J,1)}\left(\Omega_r\right)\right]\cdot\vec{\mathcal{J}}\left(\vec{r} \right) \\
\mathcal{T}^{mag}_{JM} &= \int\,d\vec{r} \left[j_{J}\left(qr\right)\vec{\mathcal{Y}}^{M}_{J(J,1)}\left(\Omega_r\right)\right]\cdot\vec{\mathcal{J}}\left(\vec{r} \right).\label{eq:Tmag}
\end{align}

The contraction of lepton and hadron currents then becomes
\begin{align*}
l_0J^0 - \vec{l}\cdot \vec{J} &= \sum_{J\geq 0}\sqrt{4\pi}\sqrt{2J+1}i^{J}\left[ l_0\mathcal{M}_{J0} -l_3 \mathcal{L}_{J0}\right] \\
        &+ \sum_{J \geq 1}\sum_{\lambda = \pm 1}\sqrt{2\pi}\sqrt{2J+1}i^Jl_{\lambda}\left[ \lambda \mathcal{T}^{mag}_{J,-\lambda} - \mathcal{T}^{el}_{J,-\lambda} \right].
\end{align*}
As the matrix element is now written as a function of irreducible tensor operators evaluated between good angular momentum eigenstates the Wigner-Eckart theorem can be used to write the expectation value of these operators as a product of a $3j$-symbol and a reduced matrix element involving only the radial wavefunctions~\cite{Walecka:book}.
If we compute the square of this matrix element and sum over the unobserved initial and final state quantum numbers we can identify the nuclear responses in terms of these reduced matrix elements 
\begin{equation}
R_{CC} = \sum_{J \geq 0} \sum_{J_f,J_i} \lvert \langle J_f \| \mathcal{M}_{J} \| J_i \rangle \rvert^2,
\end{equation}
\begin{equation}
R_{LL} = \sum_{J \geq 0} \sum_{J_f,J_i} \lvert \langle J_f \| \mathcal{L}_{J} \| J_i \rangle \rvert^2,
\end{equation}
\begin{equation}
R_{CL} = \sum_{J \geq 0} \sum_{J_f,J_i} 2\mathrm{Re}\left[\langle J_f \| \mathcal{M}_{J} \| J_i \rangle 
       \langle J_f \| \mathcal{L}_{J} \| J_i \rangle^*\right] , 
\end{equation}
\begin{equation}
R_{T} = \sum_{J \geq 1} \sum_{J_f,J_i} \lvert \langle J_f \| \mathcal{T}^{el}_{J} \| J_i \rangle \rvert^2
      + \lvert \langle J_f \| \mathcal{T}^{mag}_{J} \| J_i \rangle \rvert^2,
\end{equation}
\begin{equation}
R_{T^\prime} = \sum_{J \geq 1} \sum_{J_f,J_i} 2\mathrm{Re}\left[ \langle J_f \| \mathcal{T}^{el}_{J} \| J_i \rangle 
        \langle J_f \| \mathcal{T}^{mag}_{J} \| J_i \rangle^*\right] ,
\end{equation}
where $J_i$, and $J_f$ satisfy the triangle inequality for given $J$.

This result is  general and does not depend on the transition operator. 
The only assumptions made are the structure of the hadron current in Eq.~(\ref{eq:Jmu}), and that the initial and final states have good angular momentum quantum numbers.
For details on the operator used in this work we refer the reader to Ref.~\cite{Jachowicz:JPG}.

\begin{figure*}
    \centering
    \includegraphics{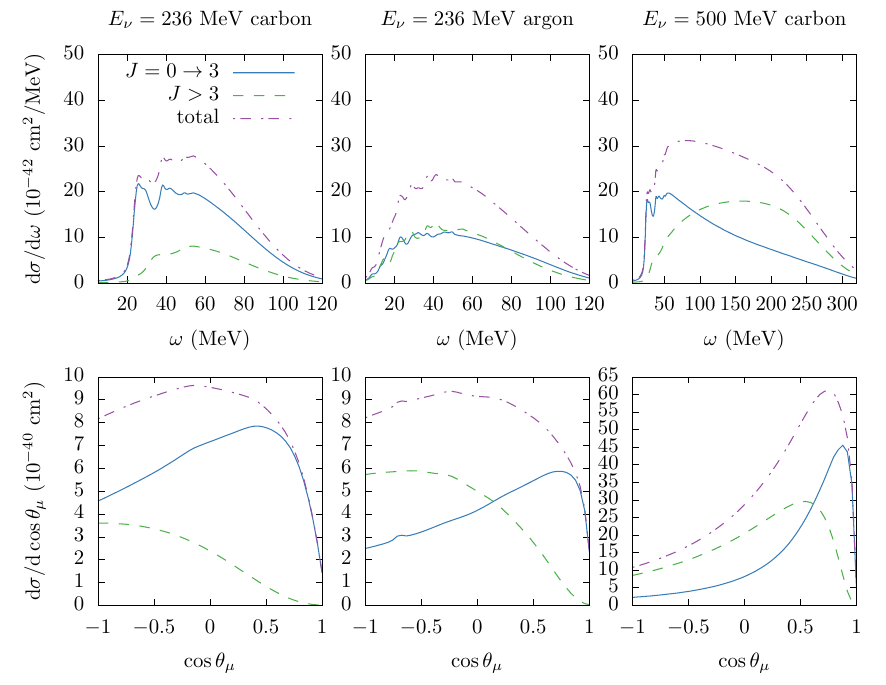}
    \vspace{-.4cm}
    \caption{Cross section per neutron for argon and carbon where the lowest order multipoles ($J \leq 3$) are shown separately.}
    \label{fig:Mpoles_A_C}
\end{figure*}

In addition to allowing for an analytical evaluation of the angular integral in the hadron current, the multipole expansion supplies information  about the dynamics of the scattering problem. In Fig.~\ref{fig:Mpoles} the relative contributions of the $J \leq 8$ multipoles to the total cross section are shown for different incoming energies and nuclei.
The $J$ in the expansion is the total angular momentum carried by the gauge boson that instigates the nuclear response. 
As the total initial nuclear angular momentum is 0, $J$ is also the angular momentum of the residual hadronic system consisting of a ``hole"-state with the angular momentum $j_h$ of the selected initial single particle state and a continuum state with angular momentum $j_p$.
The angular momentum of the hole state that can be created in the interaction is limited by the maximal available angular momentum of the initial single-particle orbitals.
For a certain $J$ the angular momentum of the outgoing nucleon $j_p$ is hence limited to $\lvert j_h - J \rvert \leq j_p \leq j_h + J$.
Thus, for low $J$ the relative weight of the lowest order partial waves for the outgoing nucleon will become more important and the process tends to be  more strongly affected by the nuclear potential and long-range correlations. 
For lower incoming energies only the smaller multipoles will contribute considerably. 
Indeed, for interactions on carbon the multipoles up to $J=3$ account for around 90 percent of the total cross section when the incoming energy is $180~\mathrm{MeV}$. For the KDAR $\nu_\mu$ this is just below $80$ percent and for the MiniBooNE peak energy this number is reduced to a mere $40$ percent.

Also the size of the nucleus plays a role, as a larger nucleus will require more partial wave contributions for the same momentum transfer. Wave functions that extend to larger $r$, will overlap more strongly with higher order spherical bessel functions $j_J(qr)$ that enter the matrix elements of Eqs.~(\ref{eq:M}-\ref{eq:Tmag}).
Another way to see this is by realizing that larger nuclei have more bound state wave functions with larger $j_h$ that will extend to larger radii, thereby allowing larger $j_p$ values for a given $J$.
This can be seen in the comparison of argon and carbon cross sections.

\begin{figure*}
    \centering
    \includegraphics{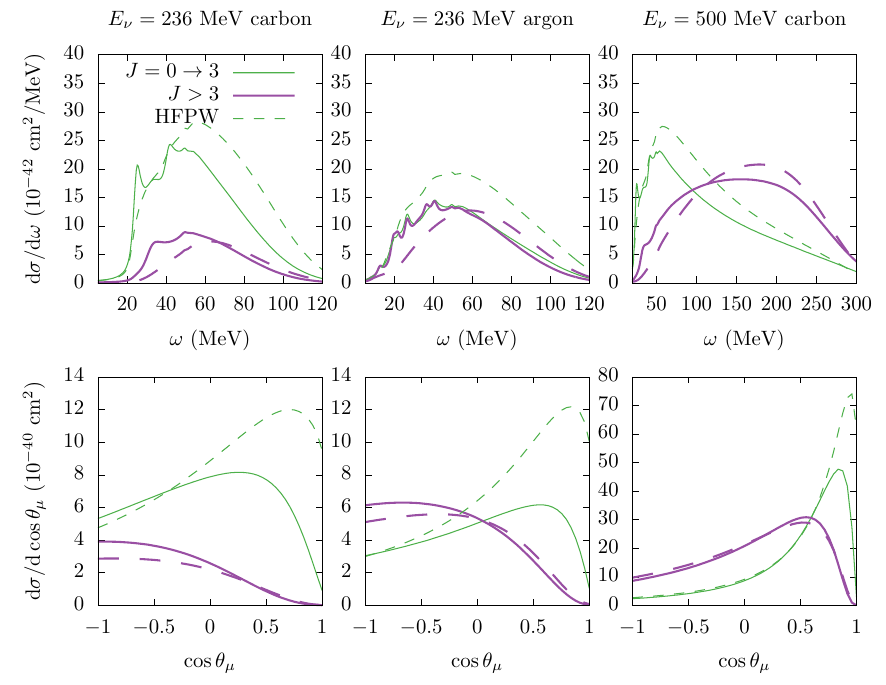}
    \vspace{-.4cm}
    \caption{ Comparison of the HF mean field results (solid lines) to the HFPW (dashed lines) in which the final state is not computed in the mean field, but is instead described by a plane wave. The contribution of the $J \leq 3$ multipoles (thin green lines) and the higher $J$ multipoles (fat purple lines) are shown separately in both cases.}
    \label{fig:Mpoles_PW}
\end{figure*}

In Fig.~\ref{fig:Mpoles_A_C} the differential cross sections in terms of energy transfer and scattering angle are shown, where the contributions from $J \leq 3$ and $J > 3$ multipoles are shown separately. 
In argon we see that higher order multipoles still carry a lot of the strength at low excitation energies, as the angular momentum of initial single particle states can be higher.

The lowest $J$ mainly contribute to the forward scattering cross section (lower $q$), while the higher multipole contributions are suppressed for forward angles. 
This shows that final-state interactions and collective effects will be more important for the forward region. This fact remains even for higher incoming energies. 
This is most clear from the results shown in Fig.~\ref{fig:Mpoles_PW}. Here the same contributions to the cross section are shown, but the HF model with final state wavefunctions obtained in the nuclear potential is compared to plane wave results (HFPW).
The operators and initial state wavefunctions in these results are exactly the same. The only difference is that the final state wavefunctions in the HFPW are not influenced by the nuclear potential.
Indeed, including final-state distortion mainly affects the lowest order partial waves.
This leads to the PW approximation overestimating the cross section especially in the forward scattering region.
As explained in Refs.~\cite{Nikolakopoulos:2019, RGJ:nucleareffects} this is mainly due to the lack of orthogonality between the initial and final state wavefunctions in the PW approach.
For a consistent description of both the cross section in terms of excitation energy and scattering angle, the consistency of initial and final state wavefunctions has to be guaranteed, especially when the relative contribution of the lowest order multipoles is large.

Summarizing, the multipole analysis makes clear that KDAR neutrinos tend to be  probing quite different aspects of the nuclear dynamics as compared to higher energy neutrinos.  As a consequence, differences in responses and cross section distributions should hence not come as a surprise.

\section{Future prospects for KDAR cross section measurements}
\label{sec:future_xsec}
There are a number of existing and planned experiments that will be able to study the KDAR $\nu_\mu$-nucleus interaction. We describe these prospects in this section.

In the near future, the MicroBooNE~\cite{uboone} and ICARUS~\cite{sbn} liquid argon time projection chamber (LArTPC) detectors, located about 102~m and 114~m from the NuMI beam absorber (discussed above), respectively, may be able to use their imaging and reconstruction capabilities to precisely isolate and measure the first monoenergetic $\nu_\mu$ interactions on $^{40}$Ar. This will likely require being able to reconstruct the few-10s-of-MeV hadronic part of the KDAR interaction products at some level, \textit{in addition} to the muon kinematics, towards a determination of the initial neutrino's direction. Direction reconstruction is important for these measurements in order to point the KDAR $\nu_\mu$ event candidates back to the absorber, given the substantial decay-in-flight neutrino background which mainly originates from the NuMI target station rather than the absorber~\cite{Spitz_xsec}. However, the pointing requirements for this purpose are very modest given the large angle between the target-station-to-detector vector (decay-in-flight background) and absorber-to-detector vector (KDAR signal). This angle is 110$^\circ$ for MicroBooNE, for example~\cite{Spitz_xsec}. Assuming production at the NuMI beam absorber at the level of 0.05~KDAR $\nu_\mu$ per proton-on-target (the NuMI target is 2.5~interaction lengths and 80\% of the power reaching the absorber is due to primary protons~\cite{NuMI_flux}, so we assume 7\% of all protons-on-target make it to the absorber 725~m downstream), corresponding to 0.7~KDAR $\nu_\mu$ per proton-on-absorber, and a KDAR $\nu_\mu$ cross section of $1.65\times10^{-39}~\mathrm{cm}^2$/neutron, consistent with the CRPA prediction described here, the 90~ton (470~ton) MicroBooNE (ICARUS) can expect 2300 (9400) events/10$^{21}$~POT. MicroBooNE has collected over 2$\times10^{21}$~POT from NuMI in both neutrino- and antineutrino-modes since data-taking began in late-2015, and ICARUS can expect 5-10$\times10^{20}$~POT/year from NuMI starting when data-taking begins in $\sim$2021. The LArTPC imaging capabilities of both MicroBooNE and ICARUS afford the ability to reconstruct $T_\mu$ and $\theta_\mu$, and therefore $q$ and $Q^2$, and aspects of the hadronic interaction as well, providing great power for differentiating between the various neutrino-nucleus interaction models discussed.

The J-PARC Materials and Life Science Facility's (MLF) Spallation Neutron Source is host to another intense source of monoenergetic KDAR $\nu_\mu$. The 3~GeV proton beam is currently providing experiments power at the 610~kW level, with a planned upgrade to 1~MW in the near future. Given the sufficient beam energy, high power, and therefore expected large KDAR signal and low decay-in-flight background, owing to the target being surrounded by concrete and iron, the J-PARC MLF is one of the leading facilities in the world to study KDAR neutrinos. The recently constructed JSNS$^2$ experiment at J-PARC MLF, a 50~ton (17~ton fiducial volume) gadolinium-loaded liquid scintillator (LS) neutrino detector located 24~m from the target, plans to make precise measurements of KDAR neutrinos using an estimated 30,000-50,000~KDAR $\nu_\mu$ CC events on carbon collected over a 3~year data taking run~\cite{Harada:2013yaa, Ajimura:2017fld}. Given the distinct double coincidence signature of these events in the light-based JSNS$^2$ detector [(1) prompt muon and hadronic-component light signal, followed by (2) the muon decay electron light signal a few-$\mu s$ later], the low beam-based background, and the $\sim10^{-5}$ beam duty factor, the signal-to-background ratio is expected to be highly favorable for a number of KDAR differential cross section measurements~\cite{Ajimura:2017fld}. In particular, the gadolinium-loaded, LS-based detector provides the ability to precisely reconstruct visible energy, which is closely related to and can be used to approximate $T_\mu + T_p$, and the number of neutrons created in KDAR $\nu_\mu$ CC interactions, although isolating the muon is more difficult since the directional Cerenkov light information is drowned out by the isotropic scintillation light. The few percent precision in visible energy reconstruction possible with JSNS$^2$ may be able to provide a unique view of the missing energy distribution in the weak interaction, which could be quite useful for neutrino interaction modeling, for example. While not competitive in terms of statistics with measurements of $(e,e^\prime p)$, the charged-current neutrino interaction brings complications not present in electron scattering as the final state does not have the same $Z$. As a consequence, neutrino interaction models tend to use effective binding energies to account for e.g. Coulomb effects and differences in proton-neutron potentials which are usually fit to inclusive data. Precision measurements in terms of visible energy can be used to study the accuracy of this modeling in detail. At the time of writing, the JSNS$^2$ experiment has started collecting neutrino beam data with the complete detector~\cite{Park:2020vxw, Park:2020uck}.

In general, future decay-at-rest sources featuring proton beam energies in excess of $\gtrsim$2-3~GeV can also serve KDAR-based physics measurements. Two prominent examples of future high-power facilities that satisfy the proton energy requirement include the planned European Spallation Source (ESS; 5-10~MW, $2.0-2.5$~GeV)~\cite{ess} and a planned multi-MW facility upgrade at Fermilab with a dedicated decay-at-rest component~\cite{eldred}.

\section{Applications and implications for neutrino experiments}
\label{sec:experiments}
Above, we have discussed the modeling of the KDAR $\nu_\mu$ cross section, the influence of relevant nuclear effects, and prospects for future cross-section-specific measurements. 
In this section, we discuss several \textit{applications} of a KDAR $\nu_\mu$ source in direct oscillation measurements and exotic searches.
We also discuss how the constraints that a measurement of KDAR neutrinos will provide can help the analyses of current and future neutrino experimental data. As an example, below we discuss how a KDAR measurement could inform the analysis of the few-100-MeV electron-like excess found in MiniBooNE, specifically the cross section modeling in the signal region, as the relevant phase space overlaps significantly with that which is probed in KDAR $\nu_\mu$ CC interactions.  

\subsection{Electron appearance from a KDAR source}

\begin{figure}
\includegraphics[width=\columnwidth]{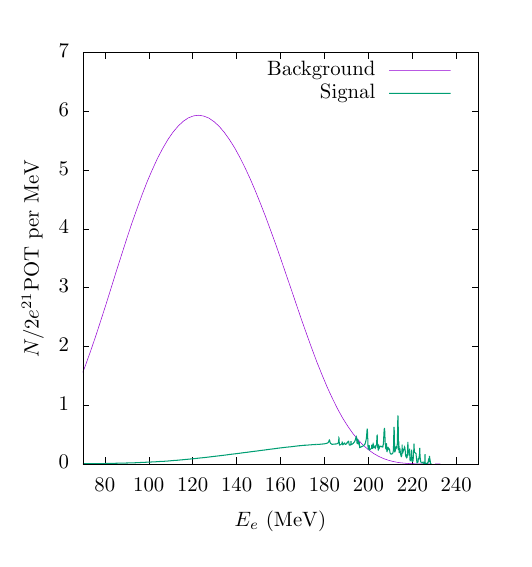}
\vspace{-1.cm}
  \caption{Expected number of $\nu_e$ ($\nu_\mu \rightarrow \nu_e$) CCQE scattering events on an argon target as a function of electron energy, where the $\nu_\mu$ is from KDAR and the background $\nu_e$ is from $K^+_{e3}$, given the assumptions of Ref.~\cite{Spitz:2012gp}. The rate for the signal and background have both been computed using the CRPA cross section.
   }
  \label{fig:el_Energy}
\end{figure}

In principle, a $236$~MeV $\nu_\mu$ KDAR source has the potential to probe the anomalous LSND and MiniBooNE signals in an electron appearance experiment ($\nu_\mu \rightarrow \nu_e$) as in Ref.~\cite{Spitz:2012gp}. 
In this section, we investigate the signal and background for such a setup with the CRPA cross sections. The envisioned setup, dedicated to creating and detecting KDAR neutrinos, consists of a 2~kton LArTPC detector at a distance of 160~m set in the opposite direction of the 8~GeV Booster Neutrino Beam at Fermilab, with all protons directed onto a large copper-based absorber. The main irreducible background that is considered comes from $\nu_e$ that are produced in decays of stopped kaons via $K^+_{e3}$ ($K^+ \rightarrow e^+ \pi^0 \nu_e$), which have an energy distribution that extends to just below $236~\mathrm{MeV}$.
In Ref.~\cite{Spitz:2012gp}, a LArTPC is proposed in order to measure the outgoing electrons and protons in $\nu_e$ CCQE events in order to reconstruct the incoming neutrino energy.
While a LArTPC in a monochromatic neutrino beam shows great potential to study the exclusive final state, the reliance on using the outgoing nucleon to reconstruct the energy leads to severe modeling complications in the kinematic regime under consideration.
For example, a model has to be able to reliably describe the missing energy distribution of argon which is a hidden variable that enters in the energy balance.
It is also well established that the plane wave impulse approximation breaks down in this kinematic regime where $q < 300~\mathrm{MeV}$. 
Therefore great care has to be taken in describing the nuclear matrix element by using a suitable (complex) potential. 
This description falls outside the scope of the present work, but instead we look at how a similar measurement could be performed by only considering the outgoing lepton kinematics.
In Fig.~\ref{fig:el_Energy} we show the expected electron energy distribution using the same assumptions made in Ref.~\cite{Spitz:2012gp}, although here no cuts are made in the phase space.
The results are very similar to what is expected by using a reconstruction method based on a simultaneous measurement of the outgoing proton, but with the advantage that the underlying missing energy distribution and specific details of the interaction beyond the inclusive cross section do not have to be known in great detail.
In such an experiment the measurement of the abundant $(\nu_\mu,\mu^- p)$ signal could be used to constrain the modeling of these nuclear structure details.
We have also checked that the reconstructed neutrino energy, as employed in accelerator-based experiments that can only measure the outgoing lepton energy, provides no clear separation between the signal and background in this case as the distribution of both the signal and background are spread out too broadly.  
It should also be clear from these considerations that an excellent energy resolution and knowledge of the background signal is necessary for such a measurement.
Lastly, it is interesting to note that the number of events of both the background and the signal obtained here match to a great extent with the results in Ref.~\cite{Spitz:2012gp} in which the electron and proton have to be measured in coincidence--one would perhaps expect that by introducing this additional complication, the magnitude of the signal (and background) would drop considerably as the proton can go undetected for many reasons.

\subsection{Informing electron appearance in MiniBooNE}
\begin{figure*}
\includegraphics{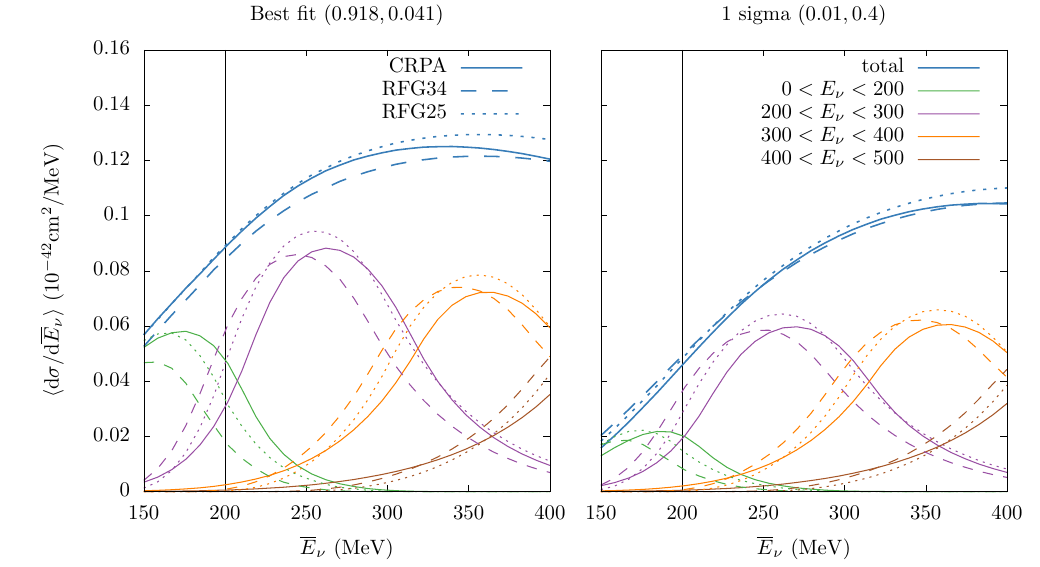}
\vspace{-.3cm}
\caption{Flux-averaged cross sections in terms of reconstructed energies for the MiniBooNE $\nu_e$ flux from $\nu_\mu \rightarrow \nu_e$ oscillations. The best-fit oscillation scenario point (left) and an example point within 1$\sigma$ of the best fit (right) are shown. The contributions from different energy bins are shown separately, with the vertical line representing the analysis threshold of $\overline{E}_\nu=200~\mathrm{MeV}$. Here the binding energy used in the definition of the reconstructed energy is taken to be $34~\mathrm{MeV}$ to match with the value used in the MiniBooNE analysis~\cite{MB:CCQE}. The dashed lines correspond to a RFG model with a binding energy of $34~\mathrm{MeV}$ while in the dotted lines $25~\mathrm{MeV}$ is used, i.e. the former consistent with the reconstructed energy and the latter using the average binding energy of the CRPA model.
All the CRPA and RFG calculations use the same value for the axial mass of $M_A=1.05~\mathrm{GeV}$.}
\label{fig:MB_Erec}
\end{figure*}
The MiniBooNE experiment has reported an anomalous excess of electron-like events, possibly indicative of $\nu_e$ appearance, in a $\nu_\mu$ beam in reconstructed energy bins of $200 - 475~\mathrm{MeV}$~\cite{MB:excess2018,MB:excess2020}.
The analysis of the excess in terms of constraining the flux and interpreting the data in any oscillation scenario relies on neutrino cross section modeling, including the ratio of the $\nu_e$ and $\nu_\mu$ cross sections in terms of real and reconstructed energies.
The reconstructed energy $\overline{E}_\nu$ is defined as 
\begin{equation}
\overline{E}_\nu = \frac{2M_n^{\prime}E_l - \left( M_n^{\prime 2} + m_l^2 - M_p^2\right) }{2\left( M_n^{\prime} - E_l + P_l\cos\theta_l\right)}.    
\end{equation}
With $M_p$ the proton and $M_n$ the neutron mass, this corresponds to the energy of a neutrino that produces a charged lepton of mass $m_l$ with energy $E_l$ and scattering angle $\theta_l$ with respect to the neutrino direction on a stationary nucleon.
Here a correction for the binding energy of the nucleon is introduced in $M_n^\prime = M_n - E_B$, with $E_B$ a fixed average binding energy. 
The cross section, and the $\nu_\mu$ to $\nu_e$ ratio, is model-dependent, especially for low-energy neutrinos where the treatment of Pauli-blocking, binding energy, and long-range correlations is crucial, and different models give varying predictions~\cite{Ankowski:PRC96, Martini:Jachowicz, Nieves:AnPh2017, Day:enu, Nikolakopoulos:2019}.

To illustrate the energy region to which the MiniBooNE analysis is sensitive we take the reported best-fit two-neutrino oscillation parameters $\sin^2 2\theta = 0.918$ and $\Delta m^2 = 0.041~\mathrm{eV}^2$ in order to characterize the flux.
Given these parameters we then obtain the $\nu_e$ flux as
\begin{equation}
\Phi_{\nu_e}\left(E_\nu\right) = \sin^22\theta \sin \left(\Delta m^2 \frac{L}{E_\nu}\right) \tilde{\Phi}_{\nu_\mu}\left(E_\nu\right),
\end{equation}
where $\tilde{\Phi}_{\nu_\mu}$ is the normalized $\nu_\mu$ flux~\cite{MB:flux} and for simplicity, we assume a single baseline of $L=540~\mathrm{m}$.\
This flux thus represents the number of $\nu_e$ per $\nu_\mu$ per unit of energy.
In Fig.~\ref{fig:MB_Erec} we show the flux-averaged cross sections at low reconstructed energy given these oscillation parameters. We also show the results for a reported allowed point within 1$\sigma$ of the best fit, which is used as an example in Ref.~\cite{MB:excess2018}  ($\sin^22\theta =0.01$, $\Delta m^2 = 0.4~\mathrm{eV}^2$).
We focus on the region $200~\mathrm{MeV} < \overline{E}_\nu < 400~\mathrm{MeV}$ as the number of observed events in this kinematic region is still significantly larger than the best-fit 3+1 oscillation scenario. Notably, this excess resides in the most forward scattering bin and for low visible energy~\cite{MB:excess2020}.
Overall, the CRPA and RFG results agree well for the total strength for both fluxes.
This happens because the strength lost in the RFG at lower neutrino energy is recovered in the next, adjacent energy bin(s).

\begin{figure}
\includegraphics[width=\columnwidth]{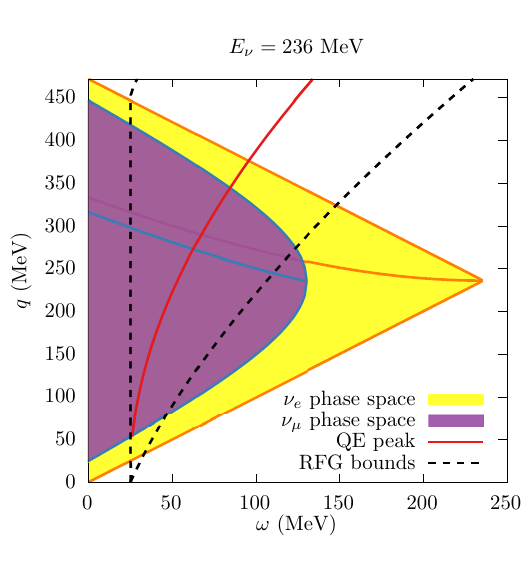}
\vspace{-.8cm}
\caption{The phase space covered in inclusive $\nu_\mu$ and $\nu_e$ charged-current scattering. Both areas are divided in forward and backward scattered charged-leptons, corresponding to the low and high $q$ region respectively. The dashed lines are the boundaries between which the RFG response with $k_F = 228~\mathrm{MeV}$ and $E_B = 25~\mathrm{MeV}$ is non-zero.}
\label{fig:PS_e_mu}
\end{figure}

In Fig.~\ref{fig:MB_Erec}, one sees that the low incoming neutrino energy regions, in particular 200-300~MeV (mainly characterized by CCQE interactions), dominate the signal in which the largest MiniBooNE data excess is found. KDAR $\nu_\mu$ cross section measurements (236~MeV) can therefore provide important constraints on the $\nu_\mu$ and $\nu_e$ CCQE cross sections in this essential region of low true and reconstructed energy. We first consider the $\nu_\mu$ cross section that in the experimental analysis is constrained by measurements of the CC $\nu_\mu$ signal in the MiniBooNE detector.
The $\nu_\mu$ flux peaks around $E_\nu = 600~\mathrm{MeV}$ with a significant high-energy tail, and the measured cross section is therefore mostly sensitive to the ``genuine'' quasi-elastic process and multinucleon emission characteristic of these energies.
The description of the $\nu_\mu$ event rate in the MiniBooNE analysis is naturally most accurate in the region where the event rate is the largest, namely for $\overline{E}$ between $0.5$ and $1.2~\mathrm{GeV}$.
Analysis of the KDAR $\nu_\mu$ cross section could thus significantly improve the modeling for lower (reconstructed) energies to which the $\nu_e$ signal is sensitive.
For example, the KDAR $\nu_\mu$ cross section can provide insight in the treatment of Pauli blocking which is empirically increased in the MiniBooNE analysis in order to better fit the data~\cite{MB:CCQE}.
Moreover, purely CCQE scattering in the low $\overline{E}$ bins can be examined without the complicating contributions from multinucleon knockout and other processes arising from higher energy $\nu_\mu$~\cite{Martini:reco, Lalakulich:reco, Nieves:unfolding, Leitner:reco}.

For the $\nu_e$ cross section, there is a significant overlap in the accessible phase space for KDAR $\nu_\mu$ scattering, as shown in Fig.~\ref{fig:PS_e_mu}. 
As previously explained, the cross section factorizes into terms that are products of, in principle well-known, leptonic prefactors and the nuclear responses.
The nuclear responses do not depend on the mass of the final state lepton but only on $\omega$ and $q$, the energy and momentum transfer to the nucleus respectively.
Therefore a comparison of the $\nu_\mu$ cross section with KDAR data in terms of $\omega$ and $q$ also provides important insight into the $\nu_e$ cross section.
A combined analysis of KDAR $\nu_\mu$ data and the available $(e,e^\prime)$ data in this kinematic region with a nuclear model that treats these processes consistently would be able to more strongly constrain the inclusive cross section in the relevant kinematic region and thus improve the understanding of the MiniBooNE anomaly.

\subsection{Other KDAR applications}
\begin{figure}
\includegraphics{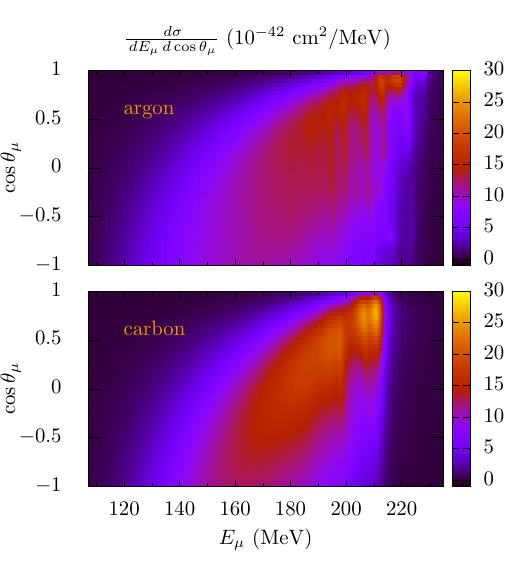}
\vspace{-.8cm}
\caption{Double differential cross section for CC scattering of a 236 MeV $\nu_\mu$ in terms of lepton kinematics. Top panel shows argon and bottom panel shown carbon cross sections per neutron.}
\label{fig:DDif_muon}
\end{figure}
In addition to the experiments poised to provide KDAR cross section measurements, including JSNS$^2$, MicroBooNE, and ICARUS (discussed above), there are a number of proposed oscillation and exotic search experiments and measurements which will rely on KDAR $\nu_\mu$ and KDAR $\nu_e$ (from $\nu_\mu \rightarrow \nu_e$ oscillations) and the predicted kinematics of the events for separating signal and background. These are presented briefly below. As usual, the sensitivity of these searches, the systematic uncertainties in the actual analyses, and the optimized design of such experiments/analyses, will be highly dependent on robust predictions for the neutrino-nucleus scattering and the resulting experimental observable signals. The CRPA predictions outlined here, in particular in terms of lepton kinematics (see Fig.~\ref{fig:DDif_muon}) can be used to reliably model and simulate KDAR events in all of these cases.

Along with the possibility of looking for KDAR $\nu_\mu \rightarrow \nu_e$ oscillations at short baseline (discussed above), there is also a proposal to search for long-baseline KDAR $\nu_\mu \rightarrow \nu_e$ oscillations in the context of ``standard" 3-neutrino oscillations~\cite{harnik}. This proposal considers the 3~GeV J-PARC Spallation Neutron Source at the MLF (also utilized by the JSNS$^2$ experiment, discussed above) for KDAR production and the future water-based Hyper-K~\cite{hyperk} as the detector, separated by 295~km. Although the $\delta_{CP}$ measurement resolution with such a configuration will not be competitive with future long-baseline experiments, given the low expected event rate from the isotropic neutrino source at long distance, the relevant $L/E$ parameter space (or rather ``point", since the single-$E=236$~MeV neutrinos will travel a known-$L=295$~km) is as-yet-unexplored. Indeed, a study of 3-$\nu$ oscillations at a single, unambiguous $L/E$ point will be unprecedented and can help rule out a number of exotic scenarios such as neutrino decay and non-standard interactions (NSI) as explanations for any possible ``expected" 3-$\nu$ oscillation signal. In general, such a measurement would be highly complementary to conventional long-baseline-based determinations of the mass hierarchy and/or $\delta_{CP}$.

In addition, there is a proposal to search for KDAR $\nu_\mu$ disappearance at short-baseline called ``KPIPE"~\cite{axani}, which will also rely on the 3~GeV J-PARC Spallation Neutron Source for KDAR neutrinos as well. This experiment will study the rate of KDAR $\nu_\mu$ events as a function of distance using a long pipe of liquid scintillator (diameter=3~m and length=120~m, with the upstream end 32~m from the neutrino source), oriented radially outward from the source. A rate measurement that deviates from the expected $1/r^2$ behavior would be an indication of new physics, perhaps coming from oscillations involving a sterile neutrino at high-$\Delta m^2$. Predictions for the kinematics of the signal KDAR $\nu_\mu$-on-carbon-induced muons, which create a a double coincidence of light [(1) from the initial muon and hadronic component flash, and (2) followed by the muon-decay-induced electron flash a few-$\mu s$ later] in KDAR events will be essential for separating signal and background given the large flux of both cosmic ray muons expected in the surface-based detector and fast neutrons originating from the beam itself. 

Along with the proposed KDAR-based experiments above, the DUNE far detector LArTPC~\cite{dune_tdr_vol4} will be sensitive to KDAR $\nu_\mu$ events originating from dark annihilation in the sun~\cite{rott1,rott2}. The idea is that \textit{if} dark matter annihilation in the Sun's core produces light quarks at substantial levels, a significant flux of stopped-pion and kaon-induced neutrinos will be created. In this case, dark-matter-induced KDAR $\nu_\mu$ (and oscillated $\nu_\mu \rightarrow \nu_e$) may be present at Earth. In separating signal from background, which mainly comes from the broad spectrum atmospheric neutrino interactions, in terms of both energy and direction, this technique is reliant on the ability to precisely reconstruct both the energy and direction of KDAR $\nu_\mu$ events using the ultra-large DUNE LArTPC. As discussed in the relevant references outlining this idea, robust kinematic predictions for KDAR neutrino-on-argon interactions are essential for estimating the search sensitivity and building up the eventual analysis.

\section{Conclusions}\label{conclusions}
Monoenergetic 236 MeV muon neutrinos from charged kaon decay-at-rest (KDAR) provide a unique opportunity to probe the neutrino-nucleus interaction with an unprecedented rigor. In the near future, a number of existing and planned experiments, e.g. MicroBooNE and ICARUS at FNAL and JSNS$^2$ at J-PARC MLF, plan to study known-energy KDAR neutrino scattering off nuclei. 
 To this end, we presented a detailed account of the cross section calculations of 236~MeV $\nu_\mu$ scattering off nuclei within a microscopic many-body theory approach.
 Our approach starts from a Hartree-Fock picture of the nuclear ground state where nucleons are bound in the potential obtained self-consistently from a Skyrme interaction. Thereby, we incorporate long-range nucleon-nucleon correlations by means of a  continuum random-phase approximation (CRPA) framework, which is particularly important for the low energy part of the accessible phase space.

First, we studied the phase-space in terms of energy and momentum transferred to the nucleus probed by the charged-current interactions of a 236~MeV $\nu_\mu$.
A significant part of this phase space lies in the transition region between low-energy nuclear excitations, sensitive to the underlying nuclear structure, and the genuine quasielastic region.
We confronted our predictions with ($e,e’$) cross section data on $^{12}$C and $^{40}$Ca in the kinematic region that overlaps with the one accessible in KDAR neutrino interactions.
We demonstrated the capabilities of the CRPA approach in describing the lower-energy (and momentum) part of the phase space, sensitive to nuclear structure details.
A satisfactory description of ($e,e’$) data on both nuclei, for the kinematics of interest for this work, validates our approach. 

Cross section predictions for the 236~MeV $\nu_\mu$ scattering off $^{12}$C and $^{40}$Ar, which are nuclear targets of interest to a number of current and future experiments aiming to utilize KDAR neutrinos are presented. 
We compared CRPA predictions with the only available KDAR measurement, a shape-only $\nu_\mu$-$^{12}$C measurement performed by the MiniBooNE collaboration, along with the predictions of a number of theoretical models and generators.
We studied Coulomb corrections with the MEMA approach, which are non-negligible at the neutrino energy of interest.
Naturally, Coulomb corrections are larger in argon than in carbon. We then presented the decomposition of the differential cross sections into longitudinal and transverse, and axial and vector components, finding that the backward cross section is dominated by the axial current. 
We also presented a decomposition of the cross section into multipoles.
Higher order multipoles constitute a significantly larger relative contribution to the argon cross section than in carbon. 

Finally, we discussed future prospects of precision KDAR cross section measurements and implications for current and future experiments that will utilize KDAR, including direct oscillation measurements and exotic searches.
In particular, we discussed how precise measurements of KDAR $\nu_\mu$-nucleus interactions could provide powerful constraints on cross section models in the energy regime where MiniBooNE observes an excess of electron-like events.
Such measurements could also help constrain the $\nu_e$-to-$\nu_\mu$ ratio at lower energies, where the mass of the lepton in the final state affects the accessible phase space considerably, and theoretical predictions are found to be model-dependent. 

The KDAR $\nu_\mu$-nucleus differential cross section predictions presented in this work can serve as an input towards providing a robust simulation of KDAR signal events for a variety of experiments.


\acknowledgments
This research was funded by the 
Research Foundation Flanders (FWO-Flanders). VP acknowledges the support from US DOE under grant DE-SC0009824. JS gratefully acknowledges support from the Department of Energy, Office of Science, under Award No. DE-SC0007859 and the Heising-Simons Foundation. 


\newpage{\pagestyle{empty}\cleardoublepage}


\end{document}